\documentclass[pra, 10pt, twocolumn, aps, superscriptaddress, showpacs]{revtex4}
\usepackage[T1]{fontenc}
\usepackage{bigints}
\usepackage{amssymb}
\usepackage{graphicx}
\usepackage{amsmath}
\usepackage{hyperref}
\usepackage{physics}
\usepackage{bm}
\usepackage{float}
\usepackage{bbm}
\usepackage{xcolor}
\usepackage{dsfont}
\usepackage{bbold}
\usepackage{dsfont}

\definecolor{lapislazuli}{rgb}{0.15, 0.38, 0.61}
\definecolor{YKblue}{rgb}{0.0, 0.18, 0.65}
\definecolor{carmine}{rgb}{0.81, 0.09, 0.13}
\definecolor{lavender}{rgb}{0.84, 0.79, 0.87}

\hypersetup{
	colorlinks=true,
	linkcolor=YKblue,
	linktoc=page,
	citecolor=blue,
	urlcolor=YKblue
}

\begin{document}
	
\title{Supersolid light in a semiconductor microcavity}

\author{J. L. Figueiredo}
\affiliation{GoLP/Instituto de Plasmas e Fus\~{a}o Nuclear, Instituto Superior T\'{e}cnico,
Universidade de Lisboa, 1049-001 Lisboa, Portugal}

\author{J. T. Mendon\c{c}a}
\affiliation{GoLP/Instituto de Plasmas e Fus\~{a}o Nuclear, Instituto Superior T\'{e}cnico,
Universidade de Lisboa, 1049-001 Lisboa, Portugal}

\author{H. Ter\c{c}as}
\affiliation{Instituto Superior de Engenharia de Lisboa, Instituto Politécnico de Lisboa, Rua Conselheiro Emídio Navarro, 1959-0077 Lisboa, Portugal}
\affiliation{GoLP/Instituto de Plasmas e Fus\~{a}o Nuclear, Instituto Superior T\'{e}cnico,
Universidade de Lisboa, 1049-001 Lisboa, Portugal}

\begin{abstract}

Supersolidity -- the simultaneous superfluid flow and crystalline order -- has been realized in quantum atomic systems but remains unexplored in photonic platforms operating at weak light-matter coupling. We predict a supersolid phase of light in a plasma-filled optical microcavity, where photons acquire effective mass and interact via nonlocal, plasma-mediated nonlinearities. By deriving a Gross--Pitaevskii equation with a tunable photon-photon interaction kernel, we show that the light field can spontaneously crystallize into a supersolid lattice via modulational instability. Crucially, this supersolid arises from a weak photon--electron coupling enabled by virtual electronic transitions, and it does not require hybrid polariton formation. Using doped semiconductor microcavities, we identify accessible conditions (electron densities $\sim 10^{10}- 10^{11}~\mathrm{cm}^{-2}$ and optical intensities $\sim 10^{3}-10^{5}~\mathrm{W/cm}^{2}$) for experimental realization. This work establishes plasmonic cavities as a platform for correlated photonic matter with emergent quantum order.

\end{abstract}
\maketitle

\textit{Introduction}--- The realization of Bose--Einstein condensation (BEC) of photons in optical microcavities has opened a new frontier in quantum optics, with bosonic statistics and optical confinement combining to yield macroscopic coherence~\cite{PhysRevA.89.033862}. Pivotal experiments with dye-filled~\cite{klaers2010bose,Klaers2010} and semiconductor~\cite{Schofield2024,pieczarka2024bose} microcavities demonstrated that two-dimensional photon gases can thermalize and macroscopically occupy the cavity ground state when mirror curvature imparts an effective mass. These achievements established a close analogy with massive bosons and laid the foundation for subsequent explorations of quantum-fluid phenomena in light~\cite{carusotto2013quantum}.

When effective photon--photon interactions are present, in consequence of interplay with matter, optical fields can display many-body behavior. In non-condensed optical media, structures such as photonic solitons~\cite{fleischer2003observation}, vortex--antivortex pairs~\cite{wan2022toroidal}, and pattern formation~\cite{whittaker2017polariton,kiselev2025symmetry} are well established. By the same reasoning, similar physics should arise in condensed-light systems once interactions are properly engineered, leveraging macroscopic phase coherence to realize genuine quantum phases of light.

The nature of photon interactions inside the optical cavity depends on the type of host medium, and the atom-mediated case has been analyzed in depth~\cite{PhysRevA.69.063816,yang2023control,heuck2020photon,PhysRevLett.107.133602}. 
While thermo-optic~\cite{stein2023exact, stein2023hartree, calvanese2022nonlocality, de2014phase, van2014interaction, maximov2024bose} and polaritonic~\cite{nigro2025supersolidity,trypogeorgos2025emerging,tian2025towards} routes to interactions in hybrid photonic condensates have also been explored, the \emph{weak-coupling} photon regime has, to the best of our knowledge, not been systematically addressed, nor has a first-principles nonlocal Gross--Pitaevskii equation for the cavity electric field been derived in this limit. In contrast to the strong coupling (polaritonic) case, in the weak-coupling regime the condensate order parameter is a genuine {\it photonic field}, with no excitonic component, and thus {\it not} featuring the usual strong coupling avoided crossing (Rabi splitting). In other words, cavity modes remain within the photon-mass limit, and any effective interaction must originate solely from virtual electronic processes encoded in the susceptibility of the medium.

In this Letter, we study the role of a two-dimensional electron gas (2DEG) in mediating effective photon--photon interactions inside microcavities. We show how the electronic response leads to nonlocal photon--photon interactions, which can destabilize the homogeneous field and trigger the spontaneous emergence of supersolid light. By integrating out the electronic degrees of freedom, we derive a driven-dissipative Gross--Pitaevskii (GP) equation for the intracavity field,
\begin{align}
i\hbar \frac{\partial \mathbf{E}(\mathbf{r})}{\partial t} &= \left[ -\frac{\hbar^2}{2M} \bm{\nabla}^2 + V_{\text{trap}}(\mathbf{r})  - i \hbar\kappa \right] \mathbf{E}(\mathbf{r})   \nonumber \\
& + \hbar \Gamma \mathbf E_p(\mathbf{r})+\int d\mathbf{r}'  g(\mathbf{r}-\mathbf{r}') |\mathbf{E}(\mathbf{r}')|^2 \mathbf{E}(\mathbf{r}),
\label{eq_GP}
\end{align}
where $M$ is the effective photon mass, $\kappa$ and $\Gamma$ the loss and pump rate, $\mathbf E_p(\mathbf{r})$ the pump field and $g(\mathbf r)$ 
the electron-induced interaction kernel, as we detail below. Unlike a local Kerr nonlinearity, $g(\mathbf r)$ acquires an oscillatory, long-range form governed by the electronic response which shares key features with dipolar or spin-orbit coupled atomic BECs \cite{PhysRevA.84.033621, khamehchi_2014, lonard_2017}, hence establishing a new route to crystallization and supersolidity in weakly coupled photonic systems.

\textit{Photon field in a plasma environment}--- Consider an optical resonator with curved mirrors of radius $R$ separated by a distance $D_0 \ll R$. Under ideal reflecting boundary conditions and sufficiently large lateral dimensions, the cavity supports a two-dimensional photon gas with an approximately quadratic dispersion relation for in-plane  momentum~\cite{RevModPhys.85.299}. In this paraxial regime \cite{PhysRevLett.133.243802}, photons acquire an effective mass $M=\hbar\pi\ell/(\overline{c} D_0)$, where $\overline{c}$ is the light speed in the medium and $\ell$ is an integer arising from the longitudinal momentum quantization. In what follows, we focus on a single longitudinal mode of the cavity.  

To generate an effective photon--photon interaction, we place a 2DEG inside the cavity. At the linear level, the plasma contributes with a refractive index shift $\sim \omega_p^2 \propto n_e$ \cite{Jackson_1998, mendoncca2017bose}, with $\omega_p$ denoting the plasma frequency and $n_e$ the electron density. However, our interest is in the nonlinear regime where two photons can interact via the plasma through excitation of virtual plasma oscillation~\cite{figueiredo2023bose,PhysRevA.110.063519}. To address the nonlinear light-matter regime, we start with the total (nonrelativistic) light-matter Hamiltonian written as $\hat H = \hat H_e + \hat H_\gamma + \hat H_\text{par.} + \hat H_\text{dia.}$, where $\hat H_e$ and $\hat H_\gamma$ are the free electron and cavity-photon Hamiltonians, and $\hat H_\text{par.}$ and $\hat H_\text{dia.}$ describe the linear (\textit{paramagnetic}) and nonlinear (\textit{diamagnetic}) light--matter interactions, respectively~\cite{cohen2024atom}. Using the appropriate expansion basis, one finds:
\begin{align}
	\hat 	H_\text{par.} &= \sum_{\mathbf k,\mathbf k',\mathbf m} \mathcal M_{\mathbf k,\mathbf k'}^{\mathbf m} (\hat a_{\mathbf m} + \hat a_{\mathbf m}^\dagger) \hat c^\dagger_{\mathbf k+\mathbf k'} \hat c_{\mathbf k'},\label{H_dia}\\
	\hat 	H_\text{dia.} &= \sum_{\mathbf k,\mathbf k',\mathbf n,\mathbf m} \mathcal I_{\mathbf k,\mathbf k'}^{\mathbf n,\mathbf m} (\hat a_{\mathbf n} + \hat a_{\mathbf n}^\dagger) (\hat a_{\mathbf m} + \hat a_{\mathbf m}^\dagger) \hat c^\dagger_{\mathbf k} \hat c_{\mathbf k'} , \label{H_par}
\end{align}
with $\mathcal M_{\mathbf k,\mathbf k'}^{\mathbf m} =  e/(m A )\bm{\mathcal A_{\mathbf m}}(\mathbf k) \cdot \hbar\mathbf{k}'$, $\mathcal I_{\mathbf k,\mathbf k'}^{\mathbf n,\mathbf m} = e^2/(2m A) \int d\mathbf r \, e^{i\mathbf r\cdot(\mathbf k' - \mathbf k)} \bm{\mathcal A}_{\mathbf n}(\mathbf r)\cdot   \bm{\mathcal A}_{\mathbf m}(\mathbf r)$, $\bm{\mathcal A}_{\mathbf m}(\mathbf k)$ denoting the Fourier transform of the vector-potential expansion coefficients $\bm{\mathcal A}_{\mathbf m}(\mathbf r)$, $e$ the elementary charge, $m$ the effective electron mass and $A$ the longitudinal area. As usual, $\hat c_{\mathbf k}^\dagger$ and $\hat c_{\mathbf k}$ denote the creation and annihilation electron operators, and $\hat a_{\mathbf m}^\dagger$ and $\hat a_{\mathbf m}$ their photonic counterpart. 

\textit{Gross--Pitaevskii equation for cavity photons}--- To capture the effect of the plasma on the photon dynamics, we derive an effective Hamiltonian for the photon field by tracing out the electronic degrees of freedom. This is valid for regimes where the nominal photon energy largely exceeds $\hbar \omega_p$ and the plasma remains weakly perturbed. In Sec.~I of the Supplementary Material (SM)~\cite{supp_mat}, we show that eliminating the electron-photon interaction to order $e^2$ yields an effective photon field Hamiltonian of the form
 \begin{equation}
 	\hat H' =  \sum_{\mathbf m} \hbar \omega_{\mathbf m} \hat a^\dagger_{\mathbf m} \hat a_{\mathbf m} + \sum_{\mathbf n,\mathbf m,\mathbf p,\mathbf q} \gamma_{\mathbf n,\mathbf m}^{\mathbf p,\mathbf q}  \hat a^\dagger_{\mathbf n} \hat a^\dagger_{\mathbf m}\hat a_{\mathbf p}\hat a_{\mathbf q}, \label{H_Eff}
 \end{equation}
where  
\begin{align}
	&\gamma_{\mathbf m,\mathbf n}^{\mathbf p,\mathbf q} = 4\sum_{\mathbf k,\mathbf k'} \mathcal I_{\mathbf k,\mathbf k'}^{\mathbf n,\mathbf q} \mathcal I_{\mathbf k',\mathbf k}^{\mathbf m,\mathbf p} \frac{f_{\mathbf k'}- f_{\mathbf k}}{\varepsilon_{\mathbf k'} - \varepsilon_{\mathbf k} }  ,  \label{gamma_final}
\end{align}
denotes the amplitude of two-photon events $(\mathbf p,\mathbf q) \rightarrow (\mathbf m,\mathbf n)$ mediated by virtual electronic transitions. The summations above involve the electron distribution function $f_{\mathbf k}$ stemming from the Fermi statistics, while the energy denominator contains the electron dispersion $\varepsilon_{\mathbf k} = \hbar^2 \mathbf k^2/(2m)$.

In the limit of large photon-occupation numbers, $\langle \hat a_{\mathbf n}^\dagger  \hat a_{\mathbf n}\rangle \gg 1$, the light field can be approximated by a coherent state, allowing a mean-field treatment. In this regime, the quantum operators are replaced by complex amplitudes and the dynamics of the electric field is governed by the nonlocal GP equation~\eqref{eq_GP} (for a detailed derivation see Sec.~II of~\cite{supp_mat}). 

The most significant term of Eq.~\eqref{eq_GP} is the interacting kernel $g(\mathbf r)$, which determines the nature of photon-photon interactions from the electronic distribution. By moving to momentum space, the transformed kernel $g(\mathbf k) = \int d\mathbf r \ e^{-i\mathbf k\cdot \mathbf r} g(\mathbf r)$ simply reads 
\begin{equation}
	g(\mathbf k) = -\Bigg(\frac{ e^4 \hbar}{\epsilon_0 D_0 \omega_0^3 m^2} \Bigg)\chi(\mathbf k,\omega = 0), \label{g_k}
\end{equation}  
where $\chi(\mathbf k,\omega)$ is the Lindhard response function,
\begin{equation}
   \chi(\mathbf k,\omega)  = \frac{1}{A} \sum_{\mathbf k'} \frac{f_{\mathbf k' - \mathbf k} - f_{\mathbf k' }}{ \varepsilon_{\mathbf k' - \mathbf k} - \varepsilon_{\mathbf k' } + \hbar \omega + i\delta^+}, \label{Lindhard}
\end{equation}
which determines the polarizability of a Fermi gas. Equation~(\ref{g_k}) quantifies the effective photon--photon interaction by capturing how the electronic polarization induced by one photon controls the interaction with a second one.

For nondegenerate electrons described by a Boltzmann distribution, the interaction simplifies to a repulsive Kerr contact potential, $g(\mathbf{r}) \propto \delta(\mathbf{r})$, recovering the classical limit studied in Refs.~\cite{calvanese2022nonlocality,nyman2014interactions,calvanese2014bose}. In contrast, for a degenerate electron gas, $f_{\mathbf{k}} = 2\Theta(k_F - k)$, with $k_F = (2\pi n)^{1/2}$ the Fermi wavevector and $\Theta(x)$ the Heaviside-step function. This leads to a nonlocal interaction of the form $g(r) \sim  G(k_F r)$, with
\begin{equation}
	G(r) = J_0(r)Y_0(r) + J_1(r)Y_1(r) , \label{G(r)}
\end{equation}
and having assymptotic behavior $G(r) \sim -\sin(2r)/r^2$ for $r \gg 1$. Here, $J_n(x)$ and $Y_n(x)$ are Bessel functions of the first and second kind, respectively. The nonlocal and oscillatory behavior contrasts with the local nonlinearity displayed in conventional Kerr media. Instead, it features spatially extended regions where the effective interaction alternates in sign (attractive vs. repulsive), bearing similarities with dipolar interactions in atomic BECs responsible for the formation of supersolid matter states~\cite{PhysRevA.108.033316,mussuperS}.

\begin{figure*}
\centering
\includegraphics[width=\textwidth]{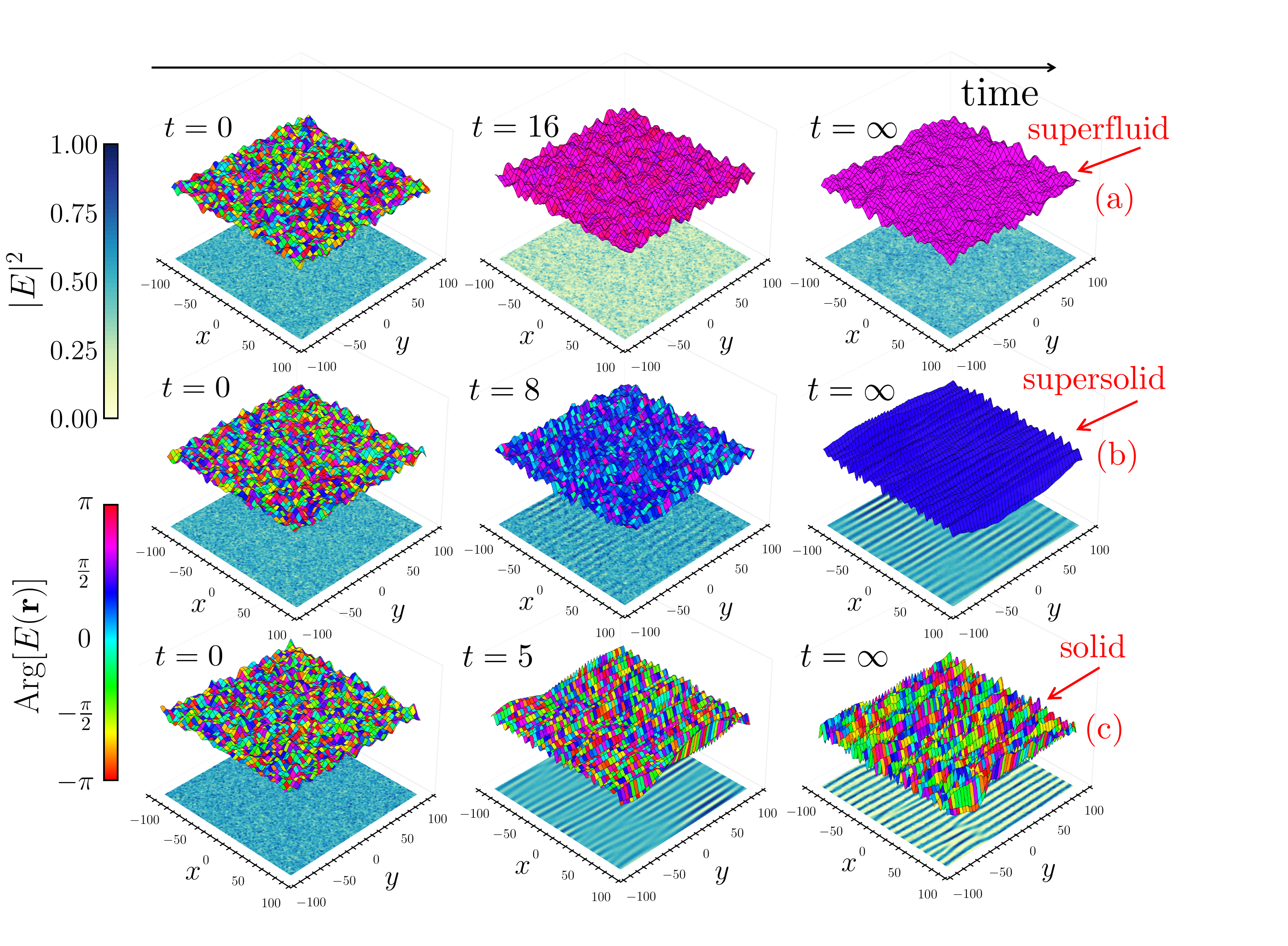}
\caption{Time evolution of the intracavity normalized field amplitude and phase for driven--dissipative conditions. The phase is indicated by te color of the 3D curve at each point, while the field amplitude is shown both in the $z$-coordinate and in the ground level in each plot. The top row shows a manifestation of superfluid behavior, where an initial homegenous field with large phase fluctuations is seen to recover global phase coherence while remaining homogeneous. The middle row illustrates the case of supersolid formation, where an initial homogeneous field breaks into a periodic lattice structure and develops global phase coherence. The bottom shows the case of a homogeneous initial condition that crystalizes while maintains large phase fluctuations, leading to a solid.  Times are measured given in units of $M/\hbar k_F^2$. Simulation parameters: $(\alpha, \kappa, \Gamma)$ = $(0.5, 0.1, 0.1)$, $(1, 0.1, 0.5)$ and $(10, 0.1, 0.5)$, respectively for top, middle and bottom rows, $N_0 = \Gamma^2/\kappa^2$ and $\Omega_p = \alpha g_0 N_0$ for all cases, and $g(\mathbf k)$ of a degenerate plasma with drift velocity $v_0 = 0.45 v_F$.}
\label{fig_phase_d}
\end{figure*}

\textit{Roton instability and supersolidity}--- The dynamical analogy between photons in 2DEG-filled cavities and atoms in dipolar BECs raises
a natural question of whether the cavity field can undergo an analogous evolution to a supersolid phase. To address this question, we examine the dispersion of collective excitations obtained from the linearized GP equation~\eqref{eq_GP}  (see Sec.~IV of~\cite{supp_mat} for a detailed derivation), 
\begin{equation}
    \hbar \omega(\mathbf k) = \pm \sqrt{\eta(\mathbf k)\Big[\eta(\mathbf k)  + 2N_0 g(\mathbf k)\Big]} - i\hbar\kappa, \label{dispRel}
\end{equation}
where $\eta(\mathbf k) = \hbar^2 k^2/(2M) - \hbar \Omega_p + N_0 g_0$, $g_0 \equiv g(\mathbf k = 0) = \int d\mathbf r \; g(\mathbf r)$, $\Omega_p$ being the pump frequency and $N_0 = |\mathbf E_0|^2$. The latter is fixed by the balance between pumping and losses according to 
\begin{equation}
	g_0^2 N_0^3 - 2 \hbar \Omega_p g_0 N_0^2  + \hbar^2(\Omega_p^2 + \kappa^2)N_0 - \hbar^2 \Gamma^2 |P|^2 = 0, \label{N0_eq}
\end{equation}
where $P$ is the pump amplitude. The emergence of a \textit{supersolid} requires two conditions to hold simultaneously~\cite{RevModPhys.82.1489,PhysRevA.110.053301,RevModPhys.84.759}: i) The long-wavelength mode must remain \textit{superfluid}, which means the dispersion to being acousting in the long-wavelength limit, i.e.  $\omega \sim v_s k$ as $k \to 0$. This can be achieved by choosing the pump frequency $\hbar \Omega_p = N_0 g_0$, which fixes $N_0$ via the self-consistency condition derived in Sec.~VI of~\cite{supp_mat}. In this case, the sound velocity becomes $v_s =\sqrt{N_0\,g_0/M}$, which further requires that $g_0 > 0$ to prevent phonon instability; ii) A finite-$k$ band must soften and eventually become unstable (roton instability). From Eq.~\eqref{dispRel}, exponential growth occurs when $\eta(\mathbf k)\big[\eta(\mathbf k) + 2N_0 g(\mathbf k)\big] < -\kappa^2$. In the high-reflectivity limit $\kappa\rightarrow 0$ this reduces to the roton-gap closure condition $\eta(\mathbf k_\ast) + 2N_0 g(\mathbf k_\ast) = 0$ at some roton wavenumber $\mathbf k_\ast\neq 0 $. Practically, this means $g(\mathbf k)$ must be positive near $k=0$ but negative in a finite region around $\mathbf k_\ast$ so the density modulations at wavelength $2\pi/|\mathbf k_\ast|$ grow while high-$\mathbf k$ modes remain stabilized by $\eta(\mathbf k)$. 

When the 2DEG is in isotropic equilibrium (i.e. in the absence of applied currents), Eq.~\eqref{g_k} yields a strictly positive interaction kernel $g({\bf k})$ for all momenta, which prevents instabilities to occur. Hence, to obtain a supersolid instability the stationary electron distribution must be anisotropic as given, e.g., by a displaced Fermi disk, $f_{\mathbf k} = 2 \Theta(k_F - |\mathbf k - \mathbf k_0|)$. This leads to an interaction kernel of the form (see Sec.~III of~\cite{supp_mat})
\begin{equation}
g({\bf k}) = -\left(\frac{e^{4}\hbar}{\epsilon_{0} D_0 \omega_{0}^{3} m^{2}}\right)
\chi_0\left({\bf k},\omega = -2\,{\bf k}\cdot{\bf v}_{0}\right),
\end{equation}
where $\chi_0({\bf k},\omega)$ is the Lindhard function of Eq.~\eqref{Lindhard} evaluated with a Fermi distribution at rest and $\mathbf v_0 = \hbar \mathbf k_0 / m$ is the drift velocity. Its analytical form in the $T \rightarrow 0$ limit was given in Ref.~\cite{stern}, which we adopt henceforth. The regions of negative $g(\mathbf k)$ are controlled by the direction and magnitude of the drift velocity ${\bf v}_{0}$ externally imposed. It is worth noting that while the screened interaction in real space displays Friedel oscillations even in a static plasma [$v_0=0$, see Eq.~\eqref{G(r)}], the corresponding momentum-space kernel $g(k)$ remains strictly positive. The external drift is therefore crucial for breaking the isotropy and rendering the effective interaction attractive in some regions of momentum where the roton instability can develop.

\textit{Dynamics of supersolid formation} ---  In Sec.~V of the SM~\cite{supp_mat}, we calculate the ground-state solutions of Eq.~\eqref{eq_GP} in the limit of a lossless cavity ($\kappa = \Gamma = 0$) hosting a current-driven 2DEG. As the chemical potential $\mu= N_0 g_0$ increases, the otherwise homogeneous solution breaks into a supersolid structure, which confirms that the translational symmetry is spontaneously broken. For the general driven–dissipative case, we expect self-organization into a supersolid when the effective interaction scale exceeds the roton-instability threshold, $2N_0g(k_*)+\eta(k_*)<0$, and the pump–loss balance stabilizes the resulting finite-momentum mode. In this regime, the supersolid emerges as a non-equilibrium steady state sustained by drive and dissipation.

Figure~\ref{fig_phase_d} shows three representative results of the cavity-field time evolution for different parameter regimes. In the top row, the homogeneous driven condensate remains stable while all Bogoliubov modes are damped, which occurs for effective interaction strengths $\Lambda= N_0 g(\mathbf k)$ below the roton threshold. In this superfluid regime the system relaxes to a uniform state with a single low-momentum peak in the structure factor and long-range phase coherence, which indicates that all density modulations seeded by noise decay, and the condensate fraction stays close to unity. Once $\Lambda$ exceeds the roton-instability line, a narrow band of finite momenta becomes dynamically unstable. The corresponding modes at $|\mathbf{k}_\ast|$ grow exponentially and imprint a density modulation with a well-defined wavelength selected by the nonlocal interaction kernel. For a broad intermediate window $\Lambda_{\mathrm{rot}} < \Lambda < \Lambda_c$, the nonlinear dynamics saturate in such a way that both the crystalline order parameter $S(\mathbf k_\ast)$ and first-order coherence $g^{(1)}(\mathbf r\rightarrow \infty)$ remain finite. In this supersolid regime, sharp Bragg peaks coexist with a substantial zero-momentum component and slowly decaying first-order coherence, indicating the simultaneous breaking of translational symmetry and preservation of phase coherence as shown in the middle row of Fig.~\ref{fig_phase_d}. We note that the explicit breaking of rotational symmetry by the drift velocity $\mathbf{v}_0$ restricts crystallization to a preferential direction, resulting in a one-dimensional periodic density modulation typical of a \textit{stripe supersolid} structure~\cite{li2017stripe,ho2011bose}. For stronger driving, $\Lambda \gg \Lambda_{\mathrm{rot}}$ (bottom row), the same roton modes dominate the density but the phase dynamics change qualitatively. The zero-momentum mode is depleted while long-range coherence collapses, and the field fragments into phase-disordered domains pinned to the lattice. The resulting normal solid exhibits a Bragg pattern nearly indistinguishable from that of the supersolid, but with vanishing $g^{(1)}(\mathbf r)$ for long distances -- check Sec.~VII of~\cite{supp_mat} for qualitative characterization analysis. Importantly, the linear spectrum is identical in the supersolid and normal-solid regions, and only the nonlinear driven–dissipative dynamics of Eq.~\eqref{eq_GP} determine whether the roton instability locks into a phase-coherent supersolid or relaxes into a phase-disordered crystal. This clarifies that the incoherent solid is a distinct non-equilibrium state where nonlinear scattering depletes the zero-momentum reservoir, destroying the phase rigidity inherent to the conservative ground state discussed in~\cite{supp_mat}.

\begin{figure}
\centering
\includegraphics[scale=0.26]{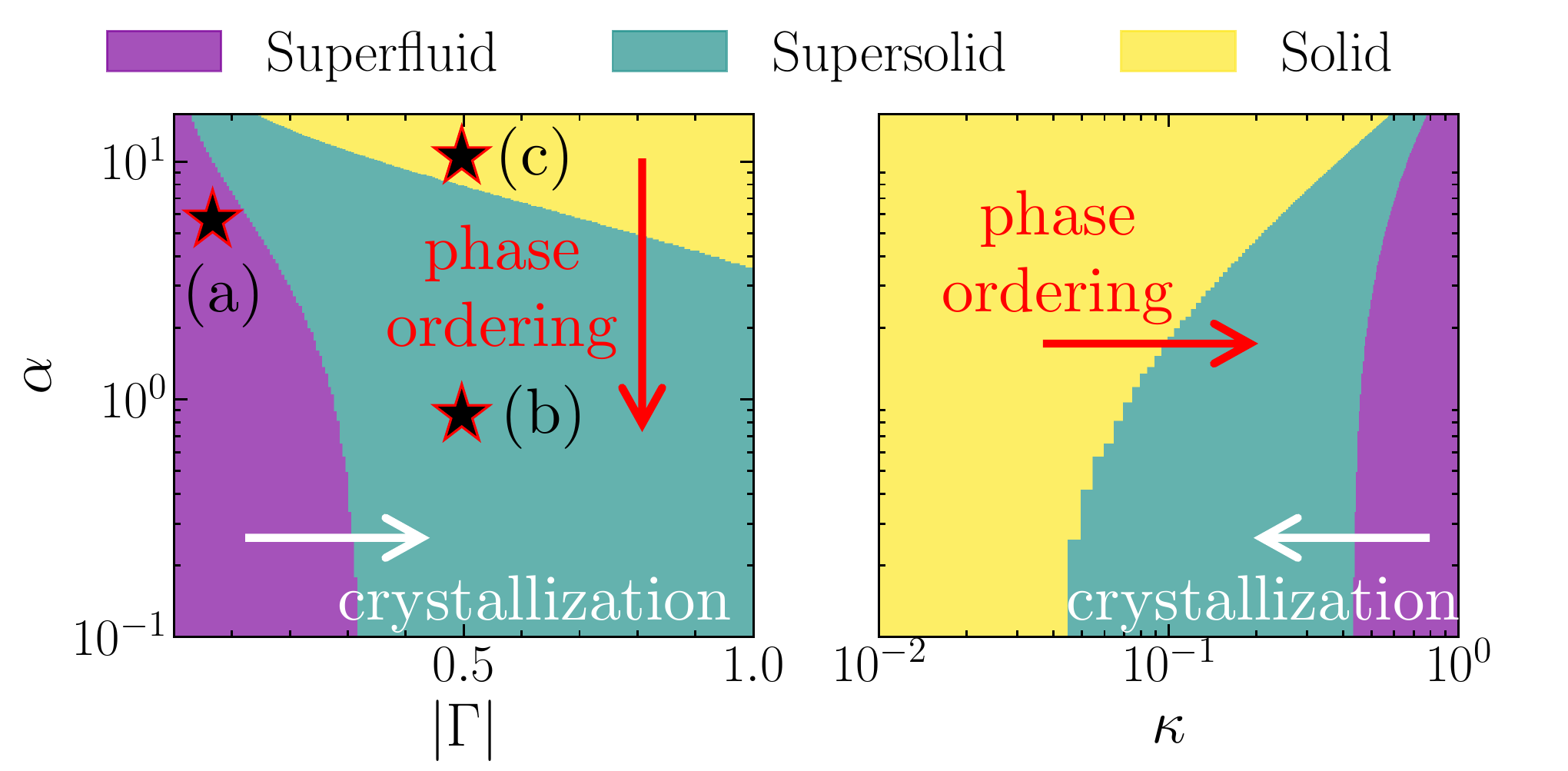}
\caption{Phase diagrams with the steady-state phase of the driven-dissipative GP equation as a function of dimensioless interaction strength $\alpha$ and (left panel) varying pump rate $|\Gamma|$ for fixed loss rate $\kappa = 0.1$, or (right panel) varying loss rate for fixed $|\Gamma| = 0.2$ (all parameters are given in dimensionless units). Each representative case of Fig.~\ref{fig_phase_d} is marked with (a) superfluid, (b) supersolid and (c) solid. The phase boundaries were extracted by evaluating two order parameters, $\lim_{x\rightarrow\infty} g^{(1)}(x)$ and $\max_{\mathbf k \neq 0} S(\mathbf k)$, where $g^{(1)}(x)$ is the first-order coherence function (averaged over $y$) and $S(\mathbf k)$ is the structure factor, and determining where they become nonzero (in practice, above a small threshold)}. 
\label{fig_diagram}
\end{figure}

These results are summarized in the phase diagrams of Fig.~\ref{fig_diagram}, indicating the phase of the steady-state solution as a function of the dimensionless interaction strength $\alpha \equiv e^4 |P|^2/(\hbar^2 \omega_0^2 \epsilon_0 m c^2 D_0 k_F^2)$, dimensionless pump rate $|\Gamma|$ and dimensionless loss rate $\kappa$. The arrows labeled crystallization and phase ordering emphasize the separation of dynamical time scales seen above in the fast selection of $\mathbf{k}_\ast$ from the unstable band set by $[\mathrm{Im}\,\omega(\mathbf{k}_\ast)]^{-1}$, followed by slower phase alignment as evidenced in Fig.~\ref{fig_phase_d}. The unstable band is centered at $|\mathbf{k}_\ast|\!\simeq\! 0.45$--$0.6$ $k_F$, implying a lattice spacing $a \!\sim\! 2\pi/|\mathbf{k}_\ast |\simeq 10$--$13$ $k_F^{-1}$. These ranges are accessible in GaAs quantum wells with typical electron densities $n_e \sim 10^{14}~\mathrm{m}^{-2}$, where $k_F^{-1} \sim 25~\mu \mathrm{m}$, yielding lattice constants $a \sim 0.4$--$0.5~\mu \mathrm{m}$.

\textit{Experimental setup}--- To realize the momentum-dependent photon--photon interaction predicted by our theory, we propose a microcavity architecture containing a single GaAs quantum well placed at an antinode of the cavity standing wave (see Sec.~VIII of~\cite{supp_mat} for scheme and additional details). The quantum well hosts a high-mobility two-dimensional electron gas with density $n_{e}\sim 10^{14}\,\mathrm{m^{-2}}$, supplied by remote Si $\delta$-doping. At these densities the electrons are in the degenerate regime, ensuring that the cavity samples the full nonlocal structure of the static Lindhard response $\chi_0(\mathbf k)$, which determines the effective interaction kernel $g(\mathbf k)$.

A small in-plane electrostatic bias applied through lateral electrodes produces a controlled drift velocity $v_{0}\simeq 0.5\,v_{F}$, displacing the electronic Fermi disk and generating the required asymmetry in $g(k_{x},k_{y})$. Because the drift is induced by a weak dc field (a few volts across a 10--30 $\mu$m mesa), the electronic distribution remains stationary and thermalized on photonic timescales ($~\sim$ fs). In this configuration the cavity field directly inherits the analytically tunable, momentum-selective interaction profile that drives the supersolid instability.

\textit{Conclusions} --- In summary, we have shown that a weakly-coupled optical microcavity containing a 2D electron gas can support a supersolid phase of light as predicted by the generalized GP equation. Above a critical threshold, the system spontaneously forms a periodic lattice of light while retaining global phase coherence, thereby realizing a supersolid state. Notably, this supersolid arises in a purely photonic regime without relying on excitonic polariton formation, distinguishing it from previously observed polariton supersolids~\cite{trypogeorgos2025emerging,zhai_2025}. We derived a Gross--Pitaevskii-type equation governing the cavity photon gas and provided analytical and numerical evidence for the supersolid transition. Importantly, our calculations identify experimentally accessible conditions for observing supersolid light. We find that electron densities of $\sim 10^{11}- 10^{12}~\mathrm{cm}^{-2}$ and optical intensities of $\sim 10^{3}-10^{5}~\mathrm{W/cm}^{2}$ are sufficient in a high-$Q$ semiconductor microcavity. Such parameters are well within the range of modern semiconductor cavity experiments, indicating that the creation of a supersolid light lattice is feasible with current technology. 

Our work provides the first evidence for a photonic supersolid in the weak coupling regime, paving the way for exploring strongly correlated phases of light in driven-dissipative systems by identifying how to manipulate the effective photon--photon coupling through plasma parameters. Beyond the single-layer implementation considered here, placing multiple parallel quantum wells inside the cavity, each sustaining a stationary in-plane current with an independently controlled direction, would enable the engineered superposition of several anisotropic electronic responses, thereby synthesizing tailored momentum-space interaction kernels $g(\mathbf{k})$. This multilayer architecture offers a route to designing light supersolids with programmable crystalline patterns, symmetry classes, and lattice geometries directly determined by the engineered structure of $g(\mathbf{k})$. Light supersolids with such engineered topology could be exploited as platforms for disorder-robust photonic routing in topological waveguides~\cite{kim2021quantum}, topological insulator laser arrays with robust phase locking~\cite{dikopoltsev2021topological}, and photonic quantum simulators of interacting and topological lattice models~\cite{ouyang2024programmable}.

\textit{Acknowledgments}--- J.~L.~F. acknowledges Funda\c{c}\~{a}o para a Ci\^{e}ncia e a Tecnologia (FCT-Portugal) through the Grant No. UI/BD/151557/2021.

\clearpage

\begin{center}
\textbf{\large Supplemental Material: Supersolid light in a semiconductor microcavity}
\end{center}
\setcounter{equation}{0}
\setcounter{figure}{0}
\setcounter{table}{0}
\setcounter{page}{1}


In this Supplementary Material we provide detailed derivations of the effective photon--photon interactions in the presence of a background plasma reservoir, and consequent generalized Gross--Pitaevskii equation for the photon field. The eigenmodes of the corresponding linearized Gross--Pitaevskii equation are determined for both the fully conservative and driven--dissipative limits, and in the former case we also find the ground state of the photon fluid via imaginary time evolution. Moreover, the conditions for photon superfluidity and supersolidity in the driven--dissipative regime are summarized, and a quantitative characterization of each phase is provided in terms of correlation functions. Finally, a possible experimental set-up aimed at detecting the supersolid of light is discussed.

\section{Photon--photon interactions in the presence of background plasma}

The interaction between cavity photons and plasma electrons leads to significant modifications of the photon dynamics. To capture these effects, one can derive an effective Hamiltonian for the photon field by integrating out the electronic degrees of freedom. In the regime where the photon energy is far from electronic resonances and the plasma remains weakly perturbed, the role of the plasma is to mediate an effective photon--photon interaction. This interaction emerges from virtual electronic transitions induced by the nonlinear light-matter potential, akin to the role of virtual phonons in mediating attractive electron--electron interactions arising in low-temperature superconductors.

 Using the rotating-wave approximation (RWA), the total electron--photon Hamiltonian can be reduced to $\hat H = \hat H_0 + \hat V$, where $\hat H_0$ denotes the plasma and photon kinetic terms,
\begin{equation}
	\hat H_0 = \sum_{\mathbf k} \varepsilon_{\mathbf k} \hat c_{\mathbf k}^\dagger \hat c_{\mathbf k} + \sum_{\mathbf n} \hbar \omega_{\mathbf n} \hat a_{\mathbf n}^\dagger \hat a_{\mathbf n}, \label{H_0}
\end{equation} 
and 
\begin{equation}
	\hat V = \sum_{\mathbf k,\mathbf k',\mathbf n,\mathbf m} \mathcal I_{\mathbf k,\mathbf k'}^{\mathbf n,\mathbf m}\hat a_{\mathbf n}^\dagger\hat a_{\mathbf m} \hat c^\dagger_{\mathbf k} \hat c_{\mathbf k'} \label{V_int}
\end{equation} 
is the number-conserving light-matter interaction. The latter corresponds to the full light-matter interacting Hamiltonian $\hat H_\text{par.} + \hat H_\text{dia.}$ upon excluding off-resonant terms.

From perturbation theory, it is possible to eliminate light--matter interactions to leading order by performing an unitary transformation $\hat U$ as described in Ref.~\cite{Kuramoto}. It is convenient to rewrite the original Hamiltonian as $\hat H = \hat H_0 + \lambda \hat V$, where $\lambda =1$ corresponds to the physical case. In the rotated Hamiltonian $\hat H' = \hat U \hat H \hat U^\dagger$ the effect of the interaction can be eliminated to first order in $\lambda$ with the choice $\hat U = \exp(i\hat S)$, where $\hat S$ is an hermitian operator that verifies
\begin{equation}
[\hat H_0 , i\hat S] = \lambda V . \label{def_S}
\end{equation}
Then, to order $\lambda^2$, the rotated Hamiltonian reads $\hat H' = \hat H_0 + \hat V_\text{eff.}$, with
\begin{equation}
\hat V_\text{eff.} =i [\hat S, \lambda \hat V] . 
\end{equation}
Using a complete basis of many-body occupation-number states $\sum_{\mathbf M} \ket{\mathbf M}\bra{\mathbf M} = \mathbbm{1}$,  Eq.~\eqref{def_S} determines the form of $\hat S$ to be  
\begin{align}
	\hat S &= \sum_{\mathbf M, \mathbf N} \frac{i \bra{\mathbf M} \lambda V \ket{\mathbf N}}{\mathcal E_{\mathbf N} - \mathcal E_\mathbf{M}} \ket{\mathbf N} \bra{\mathbf M},
\end{align}
where $\mathcal E_{\mathbf M} = \bra{\mathbf M} \hat H_0 \ket{\mathbf M}$ is the kinetic energy of the joint state $\ket{\mathbf M}$. Substituting this expression into the definition of $\hat V_\text{eff.}$ and restoring the physical condition $\lambda =1$, one finds 
\begin{align}
	\hat V_\text{eff.} = - &\sum_{\mathbf L, \mathbf M, \mathbf N } \Bigg(\frac{1}{\mathcal E_{\mathbf L} - \mathcal E_{\mathbf M} } + \frac{1}{\mathcal E_{\mathbf L} - \mathcal E_{\mathbf N} } \Bigg) \nonumber \\
	&\times  \bra{\mathbf M}  \hat V \ket{\mathbf L} \bra{\mathbf L}  \hat V \ket{\mathbf N} \ket{\mathbf M} \bra{\mathbf N} .
\end{align}

In order to get a closed effective Hamiltonian for the photon field alone, we trace out the plasma degrees of freedom by taking the expectation value of $\hat V_\text{eff.}$ over an arbitrary plasma state $\ket{\Phi}$. This requires the initial and final states to be of the form $\ket{\mathbf N} = \hat a^\dagger_{\mathbf p}\hat a^\dagger_{\mathbf q} \ket{\Phi}$ and $\ket{\mathbf M} = \hat a^\dagger_{\mathbf m}\hat a^\dagger_{\mathbf n} \ket{\Phi}$ respectively~\footnote{Lowercase letters are used to denote the single-particle states of Eq.~\eqref{H_0}, while capital letters denote occupation-number states of the joint electron + photon system}. With the help of Eq.~\eqref{V_int}, we may determine the amplitude of the process which annihilates two photons in modes $\mathbf p$ and $\mathbf q$ and creates two photons in modes $\mathbf m$ and $\mathbf n$, such that $\hat V_\text{eff.}$ becomes
\begin{align}
	\hat V_\text{eff.} =  \sum_{\mathbf m,\mathbf n,\mathbf p,\mathbf q} \gamma_{\mathbf m,\mathbf n}^{\mathbf p,\mathbf q} & \hat a^\dagger_{\mathbf m} \, \hat a^\dagger_{\mathbf n} \hat a_{\mathbf p} \hat a_{\mathbf q}. \label{V_eff}
\end{align}
After some algebra, we obtain the interaction amplitude $\gamma_{\mathbf m,\mathbf n}^{\, \mathbf p,\mathbf q}$ as
\begin{align}
	&\gamma_{\mathbf m,\mathbf n}^{\mathbf p,\mathbf q} = - 4\sum_{\mathbf k,\mathbf k'} f_{\mathbf k'}(1-f_{\mathbf k})  \mathcal I_{\mathbf k,\mathbf k'}^{\mathbf n,\mathbf q} \mathcal I_{\mathbf k',\mathbf k}^{\mathbf m,\mathbf p}  \nonumber \\
	& \times \Bigg(\frac{1}{\varepsilon_{\mathbf k} - \varepsilon_{\mathbf k'} + \hbar \omega_{\mathbf m} - \hbar \omega_{\mathbf p}} + \frac{1}{\varepsilon_{\mathbf k} - \varepsilon_{\mathbf k'} + \hbar \omega_{\mathbf q} - \hbar \omega_{\mathbf n}} \Bigg), \label{gamma_general}
\end{align}
which quantifies the strength of effective photon--photon interactions mediated by the plasma. Above, $f_{\mathbf k} = \bra{\Phi} \hat c_{\mathbf k}^\dagger \hat c_{\mathbf k} \ket{\Phi}$ is the plasma distribution function which accounts for the availability of electronic states for the virtual transitions.

Equation~\eqref{gamma_general} as it stands determines the exact photon--photon coupling up to second order in the perturbation expansion (in the present case, this means to keep light-matter interacting terms to order $e^4$). It is possible to simplify this expression by noting that the typical photon energies are much smaller than the electronic kinetic energies if one takes into account typical dimensions of the microcavity, $D_0\sim 10 - 100 \,\mu\text{m}$ and $R\sim 0.1\,$m. This provides a photon-energy spacing of the order $\Delta \sim 100\,$GHz. Assuming that $\varepsilon_{\mathbf k} - \varepsilon_{\mathbf k'}$ is of the order of the Fermi energy, we conclude that the term $\hbar \omega_{\mathbf m} - \hbar \omega_{\mathbf q}$ can be safely neglected from the denominator. Introducing this approximation into Eq.~\eqref{V_eff} and rearranging the terms, together with a change of variables $\mathbf k \leftrightarrow \mathbf k'$, leads to the simplified expression
\begin{align}
	&\gamma_{\mathbf m,\mathbf n}^{\mathbf p,\mathbf q} = 4\sum_{\mathbf k,\mathbf k'} \mathcal I_{\mathbf k,\mathbf k'}^{\mathbf n,\mathbf q} \mathcal I_{\mathbf k',\mathbf k}^{\mathbf m,\mathbf p} \frac{f_{\mathbf k'}- f_{\mathbf k}}{\varepsilon_{\mathbf k'} - \varepsilon_{\mathbf k} }  ,  \label{gamma_final}
\end{align}
which now verifies the symmetry property $\gamma_{\mathbf m,\mathbf n}^{\, \mathbf p,\mathbf q} = \gamma_{\mathbf n,\mathbf m}^{\, \mathbf q,\mathbf p}$ as expected for bosonic interactions.

\section{Derivation of the Gross-Pitaevskii equation}

In this section we derive the effective Gross--Pitaevskii (GP) equation governing the evolution of the complex electric field $\mathbf E^{(+)}(\mathbf r,t)$ under the photon--photon interaction term of Eq.~\eqref{V_eff}. 

Let $\phi_{\mathbf n}(t)\equiv \langle \hat a_{\mathbf n}\rangle$ denote the expectation value of photon annihilation operators. Equivalently, we may write $\hat a_{\mathbf n} = \phi_{\mathbf n} + \delta \hat a_{\mathbf n}$, where $\delta \hat a_{\mathbf n}$ is the quantum fluctuation verifying $\langle \delta \hat a_{\mathbf n} \rangle = 0$. The classical electric field of the photon gas $\mathbf E(\mathbf r,t)$ is fully determined by the set $\{\phi_n\}$ due to its linear expansion in photon operators. Since the GP equation in general determines a complex solution, it is convenient to separate $\mathbf E(\mathbf r,t)$ in the positive and negative frequency components, $\mathbf E(\mathbf r,t) = \mathbf E^{(+)}(\mathbf r,t) + \mathbf E^{(-)}(\mathbf r,t)$, where
\begin{equation}
	 \mathbf E^{(+)}(\mathbf r,t) = \sum_{\mathbf m} \bm{\mathcal E}_{\mathbf m}(\mathbf r) \phi_{\mathbf m}(t) \label{E_field}
\end{equation}
and $\mathbf E^{(-)} = \big[\mathbf E^{(+)}\big]^\ast$. The electric-field coefficients are related with $\bm{\mathcal A}_{\mathbf m}(\mathbf r)$ through the relation $\bm{\mathcal E}_{\mathbf m}(\mathbf r) = i \omega_{\mathbf m} \bm{\mathcal A}_{\mathbf n}(\mathbf r)$. 

To derive the equation of motion for the electric field defined in Eq.~\eqref{E_field} in an open cavity subjected to a coherent pump drive, we use the Lindblad equation,
\begin{equation}
   i\hbar  \frac{\partial}{\partial t} \hat \rho  =  [\hat H_\text{total}(t),\hat \rho] + \hat{\mathcal L}, \label{LindbladEq}
\end{equation}
where 
\begin{equation}
    \hat{\mathcal L} = \sum_{\mathbf m} \Big( 2\hbar\kappa \hat a_{\mathbf m} \hat\rho \hat a^\dagger_{\mathbf m} - \hbar \kappa
    \{ \hat a^\dagger_{\mathbf m} \hat a_{\mathbf m},\hat \rho\} \Big) 
\end{equation}
is the Lindbladian term representing cavity losses at constant rate $\kappa$. Moreover, $\hat H_\text{total}(t) = \hat H' + \hat H_\text{drive}(t)$, with $\hat H'$ the effective photon--photon Hamiltonian derived in the previous section, 
\begin{equation}
    \hat H' =\sum_{\mathbf m} \hbar \omega_{\mathbf m} \hat a^\dagger_{\mathbf m} \hat a_{\mathbf m} + \sum_{\mathbf m, \mathbf n,\mathbf p,\mathbf q} \gamma_{\mathbf m,\mathbf n}^{\mathbf p,\mathbf q} \hat a^\dagger_{\mathbf m} \hat a^\dagger_{\mathbf n} \hat a_{\mathbf p} \hat a_{\mathbf q},
\end{equation}
and 
\begin{equation}
    \hat H_\text{drive}(t) = i\hbar \sum_{\mathbf m} \Big(F_\mathbf{m}(t) \hat a^\dagger_{\mathbf m} +F_\mathbf{m}^\ast(t) \hat a_{\mathbf m} \Big)
\end{equation}
the driving Hamiltonian modeling the coherent photon injection in the cavity due to the pump field $\mathbf{E}_p(\mathbf r,t)$. The scalar functions $F_{\mathbf m}(t)$ are defined as $ F_{\mathbf m}(t)= \Gamma \int d\mathbf r \, \bm{\mathcal E}_{\mathbf m}^\ast(\mathbf r) \cdot   \mathbf{E}_p(\mathbf r,t)$, where $\Gamma$ is the pump rate.

From the definition $\phi_{\mathbf m} = \text{Tr}\big(\hat a_{\mathbf m} \hat \rho\big)$ and using Eq.~\eqref{LindbladEq}, it follows that 
\begin{align}
	i\hbar  \frac{\partial}{\partial t} \phi_{\mathbf m} &= \Big(\hbar \omega_{\mathbf m} - i\hbar \kappa \Big) \phi_{\mathbf m} + 2\sum_{\mathbf n \mathbf p \mathbf q} \gamma_{\mathbf m,\mathbf n}^{\mathbf p,\mathbf q}  \langle  \hat a_{\mathbf n}^\dagger  \hat a_{\mathbf p} \hat a_{\mathbf q}\rangle  \nonumber \\
	&+ \hbar F_{\mathbf m} . \label{dphi_dt}
\end{align}
The correlator $ \langle  \hat a_{\mathbf n}^\dagger  \hat a_{\mathbf p} \hat a_{\mathbf q}\rangle$ can be added to the system as an independent dynamical variable by calculating its corresponding equation of motion. However, by noting that $ \langle  \hat a_{\mathbf n}^\dagger  \hat a_{\mathbf p} \hat a_{\mathbf q}\rangle$ can be expanded as
\begin{align}
	\langle  \hat a_{\mathbf n}^\dagger  \hat a_{\mathbf p} \hat a_{\mathbf q}\rangle &= \phi^\ast_{\mathbf n}\phi_{\mathbf p} \phi_{\mathbf q} + \phi^\ast_{\mathbf n} \langle \delta \hat a_{\mathbf p} \delta \hat a_{\mathbf q} \rangle + \phi_{\mathbf q} \langle \delta \hat a_{\mathbf n}^\dagger \delta \hat a_{\mathbf p} \rangle \nonumber \\
    &+ \phi_{\mathbf p} \langle \delta \hat a_{\mathbf n}^\dagger \delta \hat a_{\mathbf q} \rangle +  \langle \delta \hat a_{\mathbf n}^\dagger  \delta \hat a_{\mathbf p} \delta \hat a_{\mathbf q} \rangle,
\end{align} 
we conclude that the quantum-mechanical corrections appear only at second order in the fluctuations. When the field fluctuations are small comparing to the classical value, we may neglect second-order terms and obtain a closed set of equations for the amplitudes, 
\begin{align}
	i\hbar \frac{\partial}{\partial t} \phi_{\mathbf m} &= \Big(\hbar \omega_{\mathbf m} - i\hbar \kappa \Big) \phi_{\mathbf m} +  2\sum_{\mathbf n \mathbf p \mathbf q} \gamma_{\mathbf m,\mathbf n}^{\mathbf p,\mathbf q}\phi^\ast_{\mathbf n}\phi_{\mathbf p} \phi_{\mathbf q} \nonumber \\
    &+ \hbar F_{\mathbf m} . \label{dphi_dt_2}
\end{align}

Equations~\eqref{E_field} and~\eqref{dphi_dt_2} determine the time evolution of the classical field to be
\begin{align}
	 i\hbar \frac{\partial}{\partial t} \mathbf E^{(+)}(\mathbf r) &= \sum_{\mathbf m} \hbar \Big(\omega_{\mathbf m} -i\kappa\Big) \bm{\mathcal E}_{\mathbf m}(\mathbf r) \phi_{\mathbf m} + \hbar \Gamma\mathbf{E}_p(\mathbf r) \nonumber \\
     & + 2\sum_{\mathbf m \mathbf n \mathbf p \mathbf q} \gamma_{\mathbf m,\mathbf n}^{\mathbf p,\mathbf q}  \bm{\mathcal E}_{\mathbf m}(\mathbf r)\phi^\ast_{\mathbf n}\phi_{\mathbf p} \phi_{\mathbf q},
\end{align}
where the last term on the right-hand side was obtained with the help of the orthogonality condition 
\begin{equation}
    \int d\mathbf r \, \bm{\mathcal E}_{\mathbf m}^\ast(\mathbf r) \cdot \bm{\mathcal E}_{\mathbf n}(\mathbf r) = \delta_{\mathbf m,\mathbf n}.
\end{equation}
Additionally, the first term can be easily simplified by using the definition for mode functions,
\begin{equation*}
	\hbar \omega_{\mathbf m} \bm{\mathcal E}_{\mathbf m}(\mathbf r) = \Big[-\frac{\hbar^2}{2M} \bm{\nabla}_\perp^2 + V_\text{trap}(\mathbf r_\perp) \Big]\bm{\mathcal E}_{\mathbf m}(\mathbf r) .
\end{equation*}
For the interacting term -- denoted $\mathbf U$ -- we have explicitly
\begin{widetext}
\begin{align}
	\mathbf{U}  &=  \frac{2e^4}{m^2}\frac{1}{A^2} \int d\mathbf r' \int d\mathbf r''   \sum_{\mathbf k\mathbf k'\mathbf m \mathbf n \mathbf p \mathbf q}  \frac{f_{\mathbf k'}- f_{\mathbf k}}{\varepsilon_{\mathbf k'} - \varepsilon_{\mathbf k} } \phi^\ast_{\mathbf n}\phi_{\mathbf p} \phi_{\mathbf q}e^{i(\mathbf k' - \mathbf k)\cdot (\mathbf r' - \mathbf r'')} \bm{\mathcal E}_{\mathbf m}(\mathbf r) \Big[ \bm{\mathcal A}_{\mathbf n}(\mathbf r')\cdot  \bm{\mathcal A}_{\mathbf q}(\mathbf r') \Big] \Big[ \bm{\mathcal A}_{\mathbf m} (\mathbf r'')  \cdot \bm{\mathcal A}_{\mathbf p}(\mathbf r'') \Big]. \label{U_1}
\end{align} 
By neglecting the photon energy shift to maintain consistency with the previous approximation, we may write $\mathcal E_{\mathbf m}(\mathbf r) \simeq i \omega_0  \mathcal A_{\mathbf m}(\mathbf r)$. Moreover, we have the orthogonality condition
\begin{equation}
	\sum_{\mathbf m}   \bm{\mathcal E}_{\mathbf m}(\mathbf r)  \bm{\mathcal A}_{\mathbf m}(\mathbf r') = \frac{i\hbar}{2\epsilon_0 D_0} \delta(\mathbf r - \mathbf r') \mathbf e_z.
\end{equation}
Using the above relation in Eq.~\eqref{U_1}, we get
\begin{align}
	\mathbf{U}  &=  \frac{i  e^4 \hbar}{m^2 \epsilon_0 D_0}\frac{1}{A^2} \int d\mathbf r'   \sum_{\mathbf k \mathbf k' \mathbf n \mathbf p \mathbf q}  \frac{f_{\mathbf k'}- f_{\mathbf k}}{\varepsilon_{\mathbf k'} - \varepsilon_{\mathbf k} } \phi^\ast_{\mathbf n}\phi_{\mathbf p} \phi_{\mathbf q}e^{i(\mathbf k' - \mathbf k)\cdot (\mathbf r' - \mathbf r)}  \Big[ \bm{\mathcal A}_{\mathbf n}(\mathbf r')\cdot \bm{\mathcal A}_{\mathbf q}(\mathbf r')\Big] \bm{\mathcal A}_{\mathbf p}(\mathbf r) ,\nonumber\\
	 &= - \frac{  e^4\hbar }{m^2 \epsilon_0 D_0\omega_0^3 }\frac{1}{A^2} \int d\mathbf r'   \sum_{\mathbf k \mathbf k' \mathbf n \mathbf p \mathbf q}  \frac{f_{\mathbf k'}- f_{\mathbf k}}{\varepsilon_{\mathbf k'} - \varepsilon_{\mathbf k} } \phi^\ast_{\mathbf n}\phi_{\mathbf p} \phi_{\mathbf q}e^{i(\mathbf k' - \mathbf k)\cdot (\mathbf r' - \mathbf r)}  \Big[ \bm{\mathcal E}_{\mathbf n}(\mathbf r')\cdot \bm{\mathcal E}_{\mathbf q}(\mathbf r')\Big]  \bm{\mathcal E}_{\mathbf p}(\mathbf r) , \nonumber \\
	 &= - \frac{  e^4 \hbar}{m^2 \epsilon_0 D_0\omega_0^3 }\frac{1}{A^2} \int d\mathbf r'   \sum_{\mathbf k \mathbf k' }  \frac{f_{\mathbf k'-\mathbf k} - f_{\mathbf k'}}{\varepsilon_{\mathbf k' - \mathbf k} - \varepsilon_{\mathbf k'} } e^{i\mathbf k \cdot (\mathbf r - \mathbf r')} |\mathbf E(\mathbf r')|^2 \mathbf E^{(+)}(\mathbf r) \nonumber \\
	 &= \int d\mathbf r' g(\mathbf r - \mathbf r') |\mathbf E(\mathbf r')|^2 \mathbf E^{(+)}(\mathbf r). \label{U_2}
\end{align}
\end{widetext}
In the second step we used Eq.~\eqref{E_field} and performed a change of variables $\mathbf k \rightarrow \mathbf k + \mathbf k'$ followed by $\mathbf k' \rightarrow \mathbf k' - \mathbf k$. 

Equation~\eqref{U_2} determines the kernel for nonlocal photon--photon interaction through its Fourier transform $g(\mathbf k) = \int d\mathbf r \ e^{-i\mathbf k \cdot \mathbf r} g(\mathbf r)$, which reads
\begin{equation}
	g(\mathbf k) =  - \Bigg(\frac{ e^4 \hbar }{m^2 \epsilon_0 D_0\omega_0^3 } \Bigg)\chi(\mathbf k,\omega = 0) ,  \label{g_k}
\end{equation}
and $\chi(\mathbf k,\omega)$ being the Lindhard susceptibility of the plasma,
\begin{equation}
   \chi(\mathbf k,\omega)  = \frac{1}{A} \sum_{\mathbf k'} \frac{f_{\mathbf k' - \mathbf k} - f_{\mathbf k' }}{ \varepsilon_{\mathbf k' - \mathbf k} - \varepsilon_{\mathbf k' } + \hbar \omega + i\delta^+}. \label{Lindhard}
\end{equation}
Collecting all terms, we finally obtain the GP equation~(1) of the main text, 
\begin{align}
&i\hbar \frac{\partial}{\partial t}\mathbf E^{(+)}(\mathbf r,t) = \Big[-\frac{\hbar^2}{2M} \bm{\nabla}_\perp^2 + V_\text{trap}(\mathbf r_\perp) -i\hbar \kappa \Big]\mathbf E^{(+)}(\mathbf r,t) \nonumber \\
&+ \hbar \Gamma\mathbf{E}_p(\mathbf r,t)  + \int d\mathbf r' \, g(\mathbf r-\mathbf r') \big| \mathbf E(\mathbf r',t)\big|^2 \mathbf E^{(+)}(\mathbf r,t), \label{GP}
\end{align}

\section{Photon--photon interaction potential}
Below we analyze the general features of the photon--photon interaction potential, which is determined by its Fourier transform in terms of the static Lindhard susceptibility $\chi(\mathbf k,\omega = 0)$ as defined in Eq.~\eqref{g_k}. 

We consider that the underlying electronic medium corresponds to a degenerate plasma in thermal equilibrium displaying a macroscopic current density $\mathbf j = -e n \mathbf v_0$, where $n$ is the two-dimensional electron density and $\mathbf v_0 = \hbar \mathbf k_0/m$ the drift velocity. Physically, such a current can be induced by applying an external static electric field along the cavity plane. The presence of a finite drift velocity modifies the electronic distribution function to $f_{\mathbf k} = f_\text{FD}(\varepsilon_{\mathbf k - \mathbf k_0})$, with $f_\text{FD}(x)$ the Fermi-Dirac distribution. Using this distribution function in Eq.~\eqref{Lindhard}, we obtain after straightforward algebra
\begin{equation}
	g(\mathbf k) =  - \Bigg(\frac{ e^4 \hbar }{m^2 \epsilon_0 D_0\omega_0^3 } \Bigg)\chi_0(\mathbf k,\omega = -\mathbf v_0 \cdot \mathbf k) ,  \label{g_k_0}
\end{equation}
where $\chi_0(\mathbf k,\omega)$ is the standard Lindhard susceptibility for a Fermi gas at rest. The effect of the drift velocity is to introduce a Doppler shift in the frequency argument of the susceptibility, which can be used to manipulate the photon--photon interaction potential. When $\mathbf v_0 = 0$, isotropy is restored.

The limit $T\rightarrow 0$ of $\chi_0(\mathbf k,\omega)$ was determined analytically in Ref.~\cite{stern} to be~\footnote{Note that in Ref.~\cite{stern} the susceptibility is defined with an opposite sign convention, see Eq.~(2) therein.}
\begin{align}
\chi_0(\mathbf k,\omega) = &-\frac{m}{2\pi \hbar^2} \frac{1}{z}\Big[ 2z - C_{-} \sqrt{(z-u)^2 -1}\nonumber \\
& - C_+\sqrt{(z+u)^2-1} \Big],
\end{align}
with $z = k/(2k_F)$, $u = 2 \mathbf k_0 \cdot \mathbf k/(k k_F)$, and
\begin{equation}
	C_\pm = \begin{cases}
		\text{sgn}(z \pm u) & |z \pm u| > 1, \\
		0 & |z \pm u| \leq 1.
	\end{cases}
\end{equation}
When $\mathbf k_0 = 0$, the above expression reduces to the well-known result for a two-dimensional electron gas at zero temperature~\cite{giuliani2008quantum}, 
\begin{equation}
	\chi_0(\mathbf k,0) = -\frac{m}{2\pi \hbar^2} \Bigg[ 1-\Theta\Bigg(\frac{k}{k_F}-2\Bigg)\sqrt{1-\Bigg(\frac{2k_F}{k}\Bigg)^2} \Bigg]. 
\end{equation}

\begin{figure}
\centering 
\includegraphics[scale=0.48]{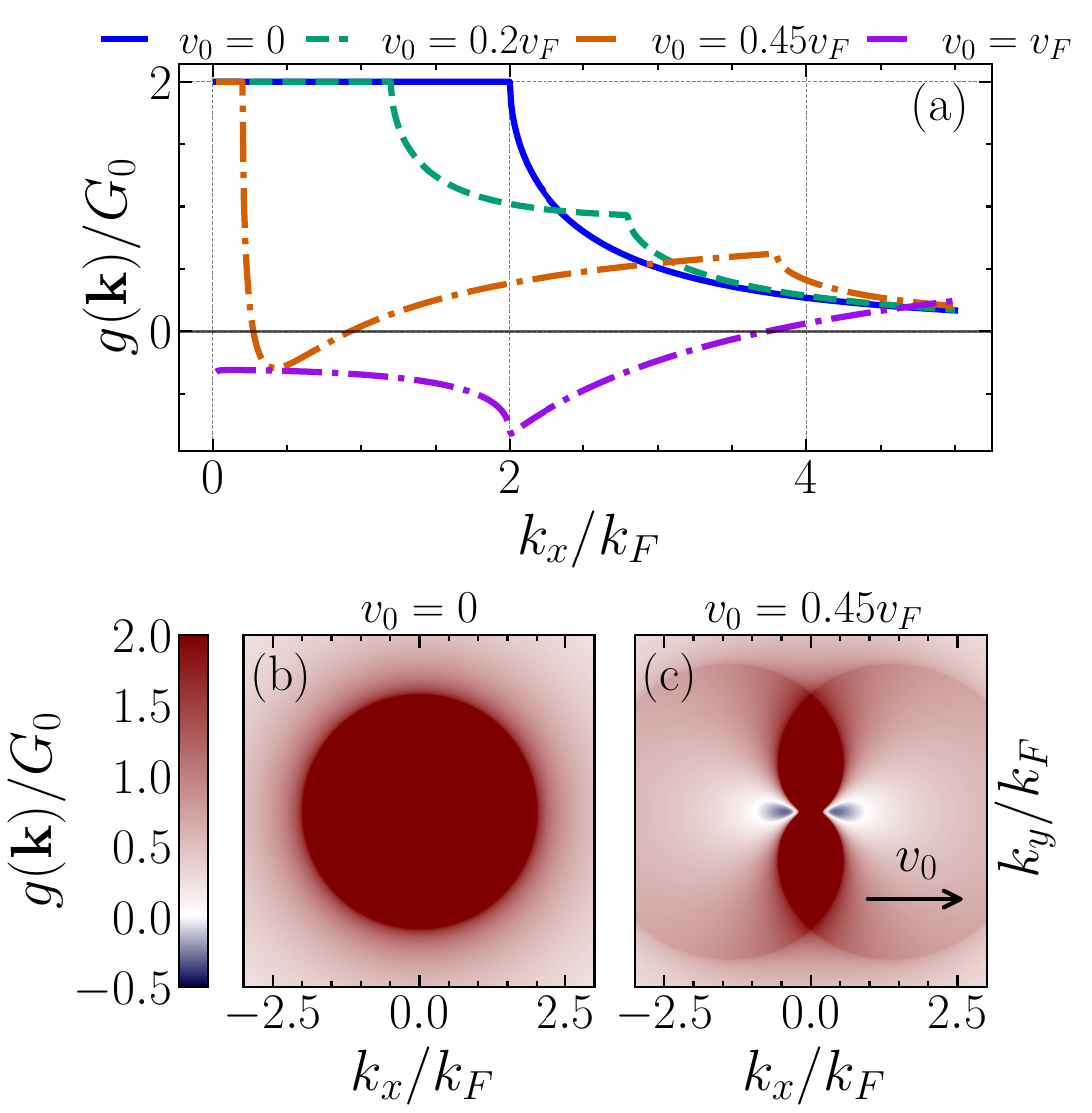}
\caption{Photon--photon interaction potential $g(\mathbf k)$ mediated by a degenerate plasma at zero temperature and moving at drift velocity $\mathbf v_0 = v_0 \mathbf e_x$. Panel (a) shows $g(k_x,0)$ for different drift velocities, while panels (b) and (c) show $g(\mathbf k)$ for both isotropic and anisotropic cases. For $v_0 = 0.45$, $g(\mathbf k)$ attains negative values in a small region along the $k_x$ direction. All velocities are given in units of the Fermi velocity, while $g(\mathbf k)$ is normalized to $G_0 = e^4 m/ (2\pi \hbar m  \epsilon_0 D_0 \omega_0^3)$.}
\label{fig_3D_panels}
\end{figure}

Figure~\ref{fig_3D_panels} shows the behavior of $g(\mathbf k)$ for different values of $\mathbf v_0 = v_0 \mathbf e_x$. As seen, the interaction potential is isotropic and repulsive when $\mathbf v_0 = 0$ [panel (b)]. However, when a finite drift velocity is considered, $g(\mathbf k)$ becomes anisotropic and attains negative values in some regions of momentum space [panel (c)]. This behavior is crucial to induce a dynamical instability in the photon fluid as described in the main text.

\section{Linear stability analysis}
In this section we derive the Bogoliubov dispersion relation of elementary excitations determined by Eq.~\eqref{GP} for a solution of the form $\mathbf E^{(+)}(\mathbf r,t) = E(\mathbf r,t) \mathbf e_z$. To this end, we assume that the trapping potential $V_\text{trap}(\mathbf r_\perp)$ is approximately constant over the region where $g(\mathbf r)$ is significant, so that we may neglect it in the stability analysis. Moreover, we consider a spatially constant pump field along the $z$ direction $\mathbf E_p(t) =  P e^{-i\Omega_pt}\mathbf e_z$, with $P = |P|e^{i\theta }$, $\theta$ being the pump phase and $\Omega_p$ the pump frequency. It is also convenient to work in dimensionless units, by scaling lengths to the electron Fermi length $r_0 = 1/k_F$, wavevectors to $k_F$, energies to $\varepsilon_0 = \hbar^2 k_F^2/M$, times to $t_0 = \hbar/\varepsilon_0$, frequencies to $1/t_0$, electric fields to the pump amplitude $|P|$, and $g(\mathbf k)$ to $G_0 = e^4/(2\pi \hbar \epsilon_0 D_0 \omega_0^3)$. From now on, we work in dimensionless units unless othewise stated. The dimensionless GP equation becomes 
\begin{align}
&i \frac{\partial}{\partial t} E (\mathbf r,t) = \Big[-\frac{1}{2} \bm{\nabla}_\perp^2-i \kappa \Big]E (\mathbf r,t) + \Gamma e^{-i\Omega_p t}  \nonumber \\
&+ \alpha \int d\mathbf r' \, g(\mathbf r-\mathbf r') \big|E(\mathbf r',t)\big|^2 E(\mathbf r,t), \label{GP}
\end{align}
where we absorbed the phase $e^{i\theta} $ in the pump rate $\Gamma \rightarrow \Gamma e^{i\theta}$ and defined the dimensionless coupling constant $\alpha$ as $\alpha = G_0 |P|^2/\varepsilon_0$.

In what follows, we treat the limit of a closed cavity $\kappa = \Gamma = 0$ and the general driven--dissipative case separately. 

\subsection{Closed cavity ($\kappa = \Gamma = 0$)}
In the fully conservative case, Eq.~\eqref{GP} admits homogeneous steady-state solutions of the form $E(\mathbf r,t) = E_0 e^{-i\mu t}$, with $\mu = \alpha N_0 g_0$, $N_0 = |E_0|^2$ and $g_0 = g(\mathbf k = 0) = \int d\mathbf r \, g(\mathbf r)$. We thus look for solutions of the form
\begin{equation}
	E(\mathbf r,t) = e^{-i\mu t}\Big[ E_0 + \delta E(\mathbf r,t)\Big]. \label{ans1}
\end{equation}
After replacing Eq.~\eqref{ans1} into the conservative version of the GP equation, neglecting second-order terms in $\delta E$, and performing a Fourier transform, we are led to the secular equation
\begin{align}
	i\frac{\partial}{\partial t} &\mqty[ \delta E(\mathbf k,t) \\  \delta E^\ast(-\mathbf k,t) ] = \overline{\overline{D}}(\mathbf k) \mqty[ \delta E(\mathbf k,t) \\  \delta E^\ast(-\mathbf k,t) ],
\end{align} 
with 
\begin{equation}
	\overline{\overline{D}}(\mathbf k) = \mqty[ \frac{k^2}{2} + \alpha N_0 g(\mathbf k) & \alpha N_0 g(\mathbf k) \\
	 - \alpha N_0 g(\mathbf k)  & -\frac{k^2}{2} - \alpha N_0 g(\mathbf k) ],
\end{equation}
the dispersive matrix whose eigenvalues determine the dispersive relation. We get for the latter
\begin{equation}
	\omega_{\pm}(\mathbf k) = \sqrt{\frac{k^2}{2}\Bigg[\frac{k^2}{2} + 2 \alpha N_0 g(\mathbf k)\Bigg]} .\label{dispRel_conservative} 
\end{equation}
When $g(\mathbf k)$ is positive everywhere, as in the case of isotropic plasma medium, the dispersion relation is purely real and the homogeneous solution is stable. In that case, we expect a superfluid behavior of the photon fluid. Contrarily, if $g(\mathbf k)$ attains negative values for some $\mathbf k$, the dispersion relation becomes imaginary in that region of momentum space, signaling a dynamical instability of the homogeneous solution towards the formation of a spatially modulated phase. This is the mechanism behind the emergence of supersolidity in the conservative limit. 

\subsection{Driven--dissipative case ($\kappa, \Gamma \neq 0$)}
When the pumping is considered, the GP equation is no longer homogeneous due to the time-dependent term $\Gamma e^{-i\Omega_p t}$. We may eliminate the explicit time dependence by moving to a frame rotating at the pump frequency through the transformation $E(\mathbf r,t) \rightarrow  E(\mathbf r,t) e^{-i\Omega_p t}$. The GP equation in the rotating frame reads
\begin{align}
&i \frac{\partial}{\partial t} E (\mathbf r,t) = \Big[-\frac{1}{2} \bm{\nabla}_\perp^2-i \kappa - \Omega_p \Big]E (\mathbf r,t) + \Gamma  \nonumber \\
&+ \alpha \int d\mathbf r' \, g(\mathbf r-\mathbf r') \big|E(\mathbf r',t)\big|^2 E(\mathbf r,t). \label{GP}
\end{align}
In the rotated frame, we use the ansatz $E(\mathbf r,t) = E_0 + \delta E(\mathbf r,t)$, where $E_0$ is now a time-independent homogeneous solution satisfying the equation
\begin{equation}
	\Big(-\Omega_p - i\kappa + \alpha |E_0|^2 g_0 \Big) E_0 + \Gamma = 0. \label{cubic}
\end{equation}
Taking the modulus squared of Eq.~\eqref{cubic}, we obtain the followin cubic equation for the steady-state photon number $N_0 = |E_0|^2$,
\begin{equation}
	\alpha^2 g_0^2 N_0^3 - 2  \Omega_p\alpha  g_0 N_0^2  + (\Omega_p^2 + \kappa^2)N_0 - |\Gamma|^2 = 0, \label{cubic_N0}
\end{equation}
which corresponds to the dimensionless version of Eq.~(10) in the main text.
The remaining linear terms in $\delta E$ lead to the Bogoliubov--de Gennes equation after a Fourier transformation,
\begin{align}
	i\frac{\partial}{\partial t} &\mqty[ \delta E(\mathbf k,t) \\  \delta E^\ast(-\mathbf k,t) ] = \overline{\overline{L}}(\mathbf k) \mqty[ \delta E(\mathbf k,t) \\  \delta E^\ast(-\mathbf k,t) ],
\end{align}
with dispersive matrix 
\begin{equation}
	\overline{\overline{L}}(\mathbf k) = \mqty[\eta(\mathbf k) + \alpha N_0 g_0 - i\kappa & \alpha N_0 g(\mathbf k) \\
	 - \alpha N_0 g(\mathbf k)  & - \eta(\mathbf k) - \alpha N_0 g_0 - i\kappa  ],
\end{equation}
and $\eta(\mathbf k) = k^2/2 - \Omega_p + \alpha N_0 g_0$. The eigenvalues of $\overline{\overline{L}}(\mathbf k)$ correspond the Bogoliubov dispersion relation in the driven--dissipative case, and read
\begin{equation}
     \omega_\pm (\mathbf k) = \pm \sqrt{\eta(\mathbf k) \Big[\eta(\mathbf k)  + 2\alpha N_0 g(\mathbf k)\Big]} - i\kappa. \label{dispRel}
\end{equation} 

\section{Ground-state solutions of the conservative GP equation}

In the conservative limit of a closed cavity, the GP equation admits stationary states that minimize the total energy. These ground states can be obtained efficiently via imaginary-time evolution. Here we use this approach to determine the spatial profile of the condensate as a function of the interaction strength $\alpha$ and of the anisotropy encoded in the plasma-mediated kernel $g({\bf k})$. For repulsive long-wavelength interactions, the ground state is uniform, corresponding to a superfluid with phase rigidity. When $g({\bf k})$ develops sufficiently negative components at finite wavevectors, the lowest-energy state acquires a periodic density modulation while retaining global phase coherence, realizing a supersolid. In the weak-interaction limit, the kinetic term dominates and the ground state approaches a Gaussian profile. The numerical results agree with the linear stability analysis discussed above.

\begin{figure*}
\centering 
\includegraphics[scale=0.55]{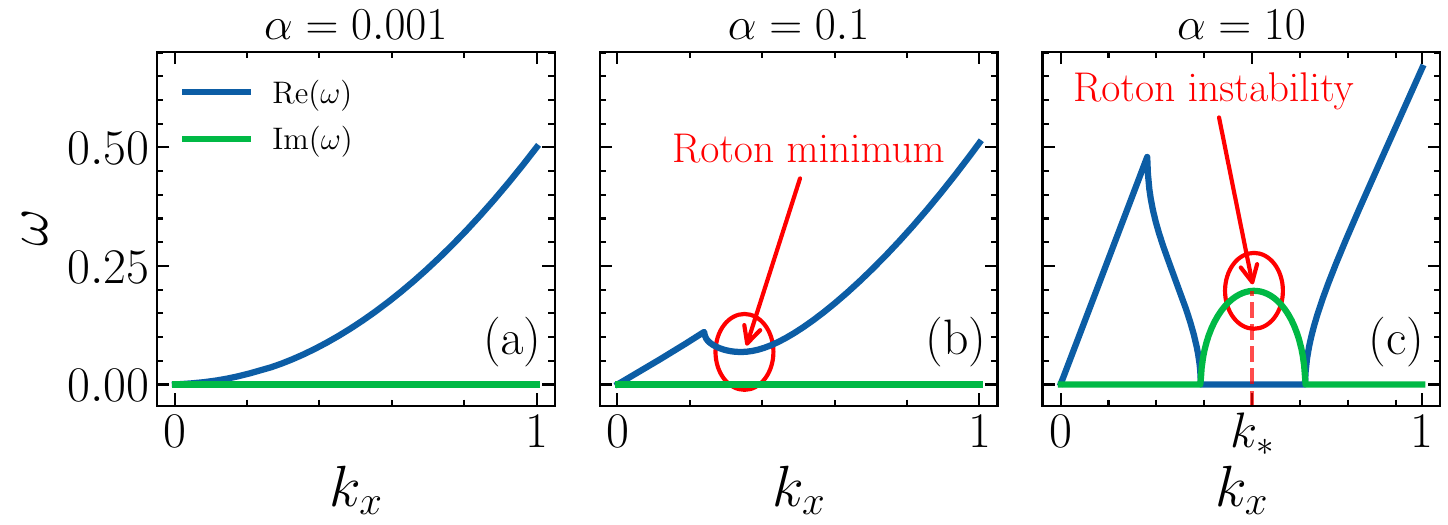}
\caption{Dispersion relation of Eq.~\eqref{dispRel_conservative} for $v_0 = 0.45 v_F$ and several values of the coupling constant along $k_y = 0$. The spectrum goes from quadratic (noninteracting limit) to a roton-like form with a finite-$k$ minimum. Instability sets in when the roton gap closes around $\mathbf k_\ast = (k_\ast,0)$, driven by increasing photon-photon interactions. }
\label{fig_disp_conservative_panels}
\end{figure*}

\begin{figure*}
\centering 
\includegraphics[scale=0.55]{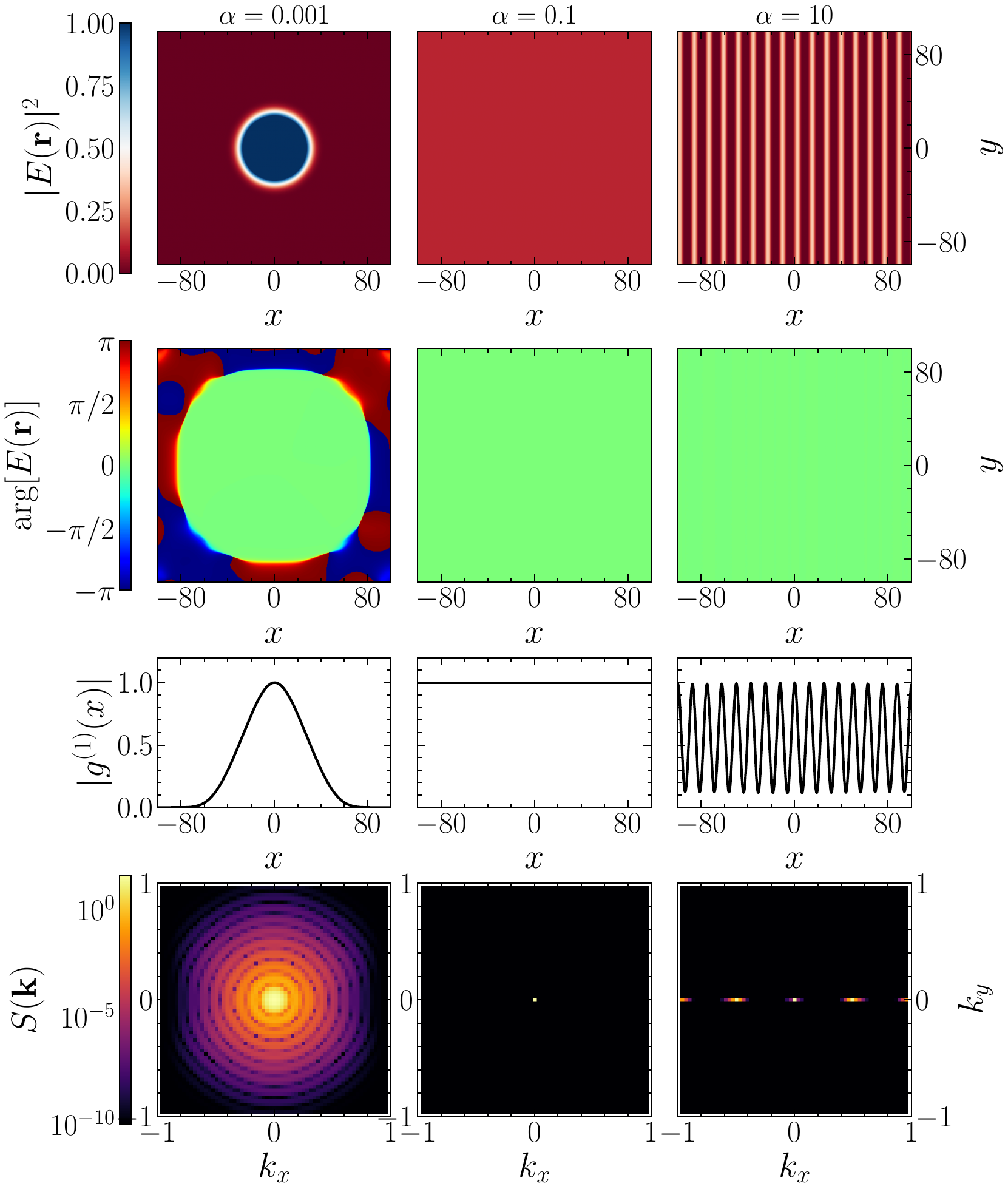}
\caption{Ground-state characterization of the conservative GP equation as a function for varying interaction strength $\alpha$ and $v_0 = 0.45 v_F$. Each column corresponds to a stationary solution obtained via imaginary-time evolution. Top row: real-space density profiles $|E(\mathbf r)|^2$, normalized to the maximum value across the three panels. Second row: phase distribution. Third row: first-order coherence $g^{(1)}(x) = \int dy \, g^{(1)}(\mathbf r)$. Bottom row: structure factor $S(\mathbf k)$ in log scale. For weak interactions ($\alpha=0.001$), the ground state is dominated by kinetic energy and approaches a smooth Gaussian profile. 
At intermediate coupling ($\alpha=0.1$), the condensate becomes a uniform superfluid with global phase coherence. For strong interactions ($\alpha=10$), attractive components of $g(\mathbf k)$ at finite wavevectors lead to density modulation and pronounced Bragg peaks in $S(\mathbf k)$ and oscillatory behavior of $g^{(1)}$ mantaining a finite value over all the entire volume, which signals the emergence of a supersolid.}
\label{fig_phase_characterization}
\end{figure*}

In Fig.~\ref{fig_disp_conservative_panels} we show the evolution of the Bogoliubov dispersion relation of Eq.~\eqref{dispRel_conservative} as the coupling constant $\alpha$ is increased for a fixed drift velocity $v_0 = 0.45 v_F$. The spectrum goes from quadratic (noninteracting limit) to a roton-like form with a finite-$k$ minimum. Instability sets in when the roton gap closes around $\mathbf k_\ast = (k_\ast,0)$, driven by increasing photon-photon interactions and triggering the spontaneous formation of a modulated stripe phase. Each case corresponds to a different ground-state solution, as illustrated in Fig.~\ref{fig_phase_characterization}. 

Figure~\ref{fig_phase_characterization} displays the corresponding ground-state configurations. The density, phase, first-order coherence $g_1(x)$, and structure factor $S(\mathbf k)$ provide a complete characterization of the emergent phases. We define the first-order coherence function as the normalized spatial autocorrelation of the field, $g^{(1)}(\mathbf{r}) = \int d\mathbf{r}' E^*(\mathbf{r}') E(\mathbf{r}'+\mathbf{r}) / \int d\mathbf{r}' |E(\mathbf{r}')|^2$, and the structure factor as the power spectrum of the density, $S(\mathbf{k}) = \left| \int d\mathbf{r} \, e^{-i\mathbf{k}\cdot\mathbf{r}} |E(\mathbf{r})|^2 \right|^2$. In the weak-interaction limit ($\alpha = 0.001$), $g_1(x)$ decays rapidly and $S(\mathbf k)$ becomes featureless, consistent with a structure near the single-particle ground state. In the superfluid regime ($\alpha = 0.1$), the density is uniform, the phase is coherent, and $g_1(x)$ remains close to unity. In the supersolid regime ($\alpha = 10$), $g_1(x)$ retains long-range coherence while $S(\mathbf k)$ acquires sharp Bragg peaks at the roton wavevector, signaling one-dimensional crystalline order characteristic of a stripe supersolid. The distance between density maxima in real space matches $2\pi/k_\ast$.

\section{Tri-stability and bifurcations in driven--dissipative conditions}
Unlike the conservative case, the presence of pumping and losses leads to a rich nonlinear dynamics characterized by multiple coexisting steady states as determined by Eq.~\eqref{cubic_N0}. The latter is a cubic polynomial in $N_0$ and for given parameters ($\alpha$, $\kappa$) and drive parameters ($\Omega_p$, $|\Gamma|$) it may have either one or three real solutions. 

\subsection{Bifurcation in the $(\Omega_p,|\Gamma|)$ plane}
For most parameter values Eq.~\eqref{cubic_N0} has a unique real solution (monostable regime). However, within a certain region of $(\Omega_p,|\Gamma|)$ parameter space the cubic yields three real roots for $N_{0}$. In that regime the system is multistable: typically two of the solutions are linearly stable (low- and high-$N_0$ branches) while the intermediate solution is unstable. The transition between the mono- and tri-stable regimes occurs via saddle-node bifurcations, where the cubic develops a double root. The boundary of this region is obtained by solving Eq.~\eqref{cubic_N0} simultaneously with $\partial W/\partial N_{0}=0$, where $W(N_{0})$ denotes the left-hand side of Eq.~\eqref{cubic_N0}.

Solving these two equations yields the critical pump strength at the saddle-node bifurcation,
\begin{equation}
|\Gamma|^2_{\rm crit} = \frac{4}{27 \alpha^2 g_0^2}
\Big( \Omega_p^2 + \kappa^2 \Big)^\frac{3}{2},
\label{cusp-eq}
\end{equation}
which defines a cusp-shaped boundary in the $(\Omega_p,|\Gamma|)$ plane. Inside this cusp the cubic discriminant is positive and three real roots exist, while outside the cusp only a single real root exists.

The cusp point itself (the tip of the tristable region) occurs where the two branches of Eq.~\eqref{cusp-eq} meet, corresponding to a triple root of Eq.~\eqref{cubic_N0}. This happens when $\Omega_p = \sqrt{3}\,\kappa$. At that point Eq.~\eqref{cusp-eq} gives $|\Gamma|^{2}_{\rm crit} = 32\kappa^3/(27 \alpha^2 g_0^2)$. 

To understand the structure of the solutions it is useful to analyse the extrema of $W(N_{0})$, the left-hand side of Eq.~\eqref{cubic_N0}. Setting $\partial W/\partial N_{0}=0$ gives
\begin{equation}
3 \alpha^2 g_0^2 N_{0}^2 - 4 \Omega_p \alpha g_0 N_{0} + (\Omega_p^2 + \kappa^2) = 0.
\end{equation}
which yields the values of $N_{0}$ at the saddle-node points. Substituting these back into $W(N_{0})=0$ leads directly to the condition~\eqref{cusp-eq} for the cusp boundary. In this way one can identify the lower and upper saddle-node curves (where the low/mid and mid/high branches, respectively, merge and annihilate) and the enclosed tristable region.

\begin{figure}
\centering 
\includegraphics[scale=0.55]{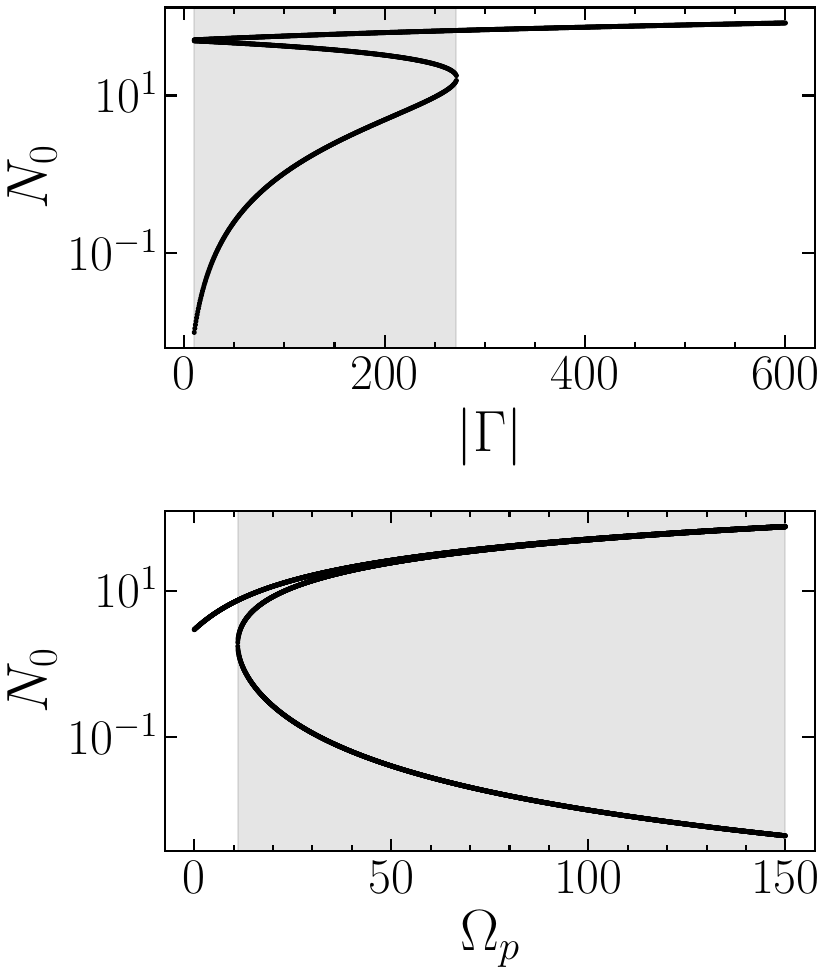}
\caption{Bifurcation diagram showing the number of real steady-state solutions $N_0$ of Eq.~\eqref{cubic_N0} as a function of pump frequency $\Omega_p$ and pump strength $|\Gamma|^2$ for fixed $\alpha = 1$, and $\kappa = 0.1$. The shaded region indicates tristability, where three real solutions coexist. The boundaries correspond to saddle-node bifurcations where two solutions merge.}
\label{fig_bifurcation}
\end{figure}

To visualise the multistability more directly, in Fig.~\ref{fig_bifurcation} we depict $N_{0}$ as a function of pump rate $|\Gamma|$ for a fixed detuning in the tristable region. Such plot exhibit the characteristic $S$-shaped response curve of a Kerr-type nonlinear resonator. The resulting curve displays three distinct branches. The turning points of the $S$-curve occur at the saddle-node bifurcations, where the intermediate branch meets the upper and lower branches. The intermediate solution is unstable, while the lower and upper branches are stable. In terms of the number of real solutions, the system is monostable at very low pump (only the low-intensity branch exists) and again at sufficiently high pump rates (only the high-intensity branch survives). In between, for an intermediate range of $|\Gamma|$, three real solutions coexist (one unstable and two stable), corresponding to effective optical bistability with hysteresis. Equivalently, we may take the same analysis in terms of the pump frequency $\Omega_p$ for fixed $|\Gamma|$, leading to a similar $S$-shaped curve also shown in Fig.~\ref{fig_bifurcation}.

\subsection{Bogoliubov stability in the tristable regime}
In a driven--dissipative system, the global phase is pinned by the pump frequency, so the otherwise Goldstone mode is not automatically gapless. At $k=0$, the Bogoliubov dispersion attains a finite value 
\begin{equation}
	\omega_\pm(0) = \pm \sqrt{(-\Omega_p + \alpha N_0 g_0)(-\Omega_p + 3 \alpha N_0 g_0)} - i\kappa.
\end{equation}
The real part vanishes if 
\begin{equation}
	\Omega_p = \alpha N_0 g_0 , \label{SF-condition}
\end{equation}
and the phase (Goldstone) mode becomes gapless, signalling superfluid behavior up to an overall damping $-\kappa$. This condition is satisfied only on the upper branch of the $S$-curve, indicating that only the high-intensity solution can exhibit superfluidity. The low- and intermediate-intensity branches have gapped phase modes and do not display superfluid behavior.

Using the superfluid condition in Eq.~\eqref{cubic_N0}, the quadratic and cubic terms cancel out, leaving a simple linear relation for the photon number on the superfluid branch, $(\Omega_p^2 + \kappa^2 ) N_0 = |\Gamma|^2$. Using again $\Omega_p = \alpha N_0 g_0$ yields 
\begin{equation}
	|\Gamma|^2 = \kappa^2 N_0 , \label{SF-pump}
\end{equation}
hence at the superfluid point, $N_0 = |\Gamma|^2/\kappa^2$ and $\Omega_p = \alpha g_0 |\Gamma|^2/\kappa^2$. Achieving a gapless mode therefore requires simultaneously tuning the pump frequency according to Eq.~\eqref{SF-condition}
and the pump power according to Eq.~\eqref{SF-pump}. At this point the dispersion is linear at small $k$ and the
photon fluid exhibits maximal phase rigidity.

Moreover, a key question is whether a supersolid phase can emerge in the presence of drive and dissipation. In a conservative system a roton instability of the superfluid indicates the transition to a supersolid: as the interaction strength increases, the Bogoliubov dispersion develops a roton minimum which eventually touches zero energy, triggering spontaneous density modulation. In the driven–-dissipative cavity a similar mechanism applies, but now the imaginary part of the Bogoliubov eigenvalues must also be considered.

The uniform condensate becomes dynamically unstable when some mode $\mathbf k$ acquires a positive growth rate $\Im[\omega(\mathbf k)] > 0$.
When the roton minimum softens sufficiently, the argument of the square root becomes negative:
\begin{equation}
\eta(\mathbf k)\,[\,\eta(\mathbf k) + 2 \alpha N_{0} g(\mathbf k)\, ] < 0 .
\end{equation}
In that regime the square root is imaginary and the two eigenvalues become purely imaginary (one decaying, one growing). When the real part of the upper branch passes through zero at the roton wavevector $\mathbf k_\ast$, one obtains
\begin{equation}
\Im[\omega_{+}(\mathbf k_\ast)]=\kappa_{\rm eff} - \kappa ,\qquad \kappa_{\rm eff} > 0 .
\end{equation}
At the instability threshold,
\begin{equation}
\Re[\omega_{+}(\mathbf k_\ast)] = 0,
\qquad
\Im[\omega_{+}(\mathbf k_\ast)] = 0.
\end{equation}
Beyond the threshold the imaginary part becomes positive, signalling exponential growth of the modulated mode. The leading condition for the onset of instability is obtained by requiring the real part of the frequency to vanish:
\begin{equation}
\eta(\mathbf k_\ast) + 2 \alpha N_{0} g(\mathbf k_\ast) = 0 .
\label{roton-zero}
\end{equation}
This condition determines the critical wavevector $\mathbf k_\ast$ and the critical interaction strength or photon number at which the uniform condensate becomes unstable. Provided the pump is strong enough to sustain the pattern against losses, the final state is a driven supersolid with both periodic density modulation and phase coherence.

\section{Drive-dissipative phase characterization}

To corroborate the analytical stability analysis, we performed numerical simulations of the driven-dissipative GP equation~\eqref{GP} on a $256\times 256$ spatial grid using a split-step Fourier method and periodic boundary conditions. The initial condition was taken as $\big[\sqrt{N_0} + \delta(\mathbf r) \big] e^{i\theta(\mathbf r)}$ where $\delta(\mathbf r) \ll \sqrt{N_0}$ is a small random noise with zero mean, $\langle \delta(\mathbf r)\rangle = 0$, and $\theta(\mathbf r)$ is a random phase. A fast Fourier transform algorithm was used to compute the nonlocal contribution via the convolution theorem. Representative cases are shown in Fig.~1 of the main text, illustrating the time evolution towards different steady states as parameters are varied. Below we provide additional figure characterization of these different phases. In order for superfluid properties to emerge, the system must be driven to the upper branch of the $S$-curve discussed above. From Eq.~\eqref{dispRel_conservative}, this requires the superfluid condition $ \Omega_p = \alpha N_0 g_0$ to be verified, so that the $k\to 0$ mode remains stable, and consequently $N_0 = |\Gamma|^2/\kappa^2$, as dictated by Eq.~\eqref{cubic_N0}. We ensured that these conditions are satisfied in all simulations presented below.

\subsection{Superfluid phase}
When the nonlinearity is small ($\alpha \lesssim 1$), no instability is present, as predicted by the linear stability analysis [Eq.~\eqref{dispRel}]. The photon fluid remains in a homogeneous superfluid state, characterized by a uniform density profile, long-range phase coherence, and a structure factor peaked at zero wavevector. 

\begin{figure}
\centering 
\hspace{-0.5cm}
\includegraphics[scale=0.3]{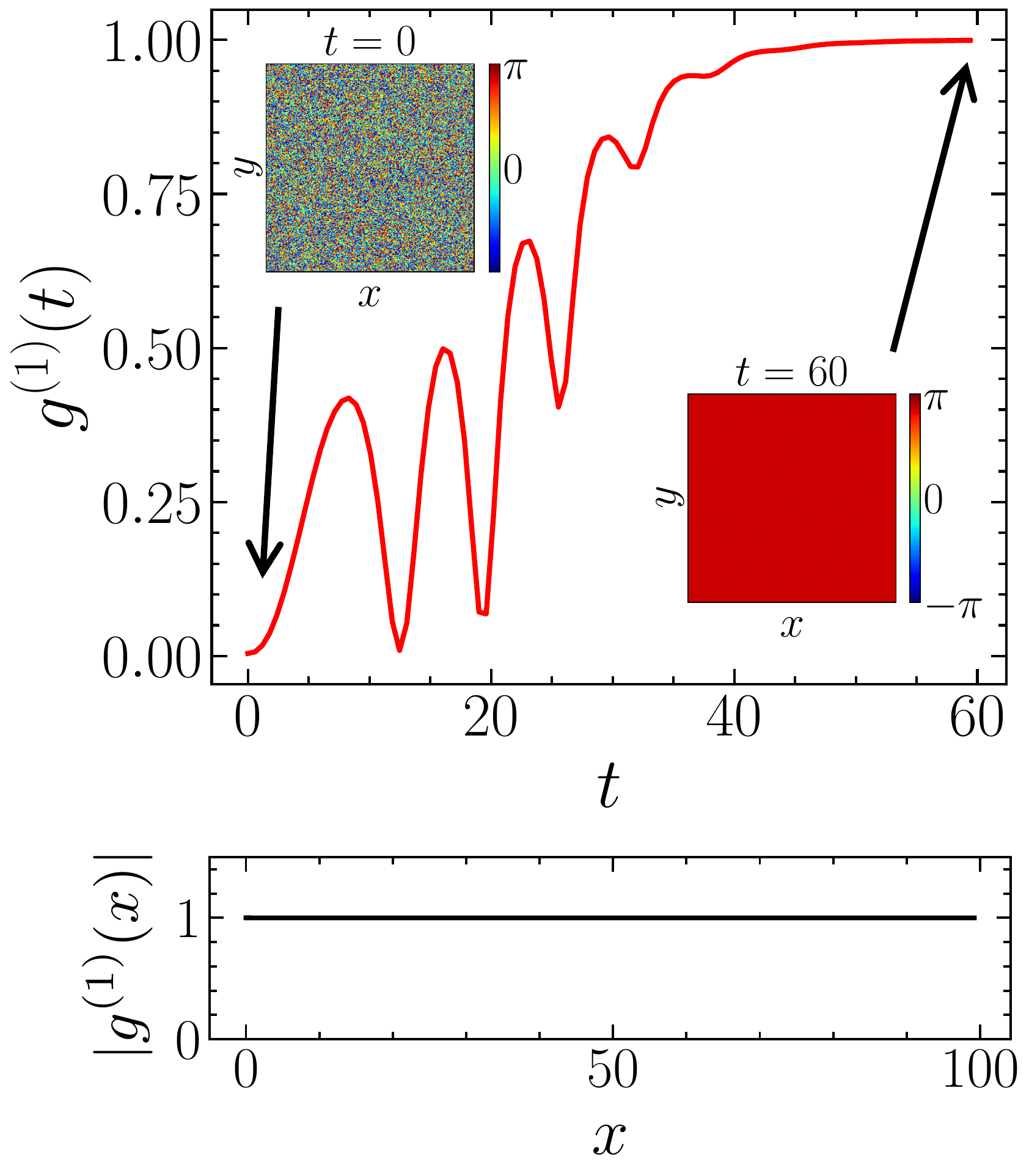}
\caption{Characterization of the superfluid phase. Top panel: long-range coherence $g^{(1)}(\mathbf r\rightarrow\infty ,t)$ as a function of time showing the build up of phase coherence from initial phase disorder. Insets show the phase of the intercavity field at the intial and final simulation times (axis are those of Fig.~1 of the main text). Bottom panel: first-order coherence function $g^{(1)}(x)$ along the $x$ direction measured at the final time $t = 60$. Simulation parameters are $\alpha = 0.5$, $\kappa = 0.1$, $\Gamma = 0.1$, $k_b = 0.45 k_F$.  }
\label{fig_SF_diagnostics}
\end{figure}

In Fig.~\ref{fig_SF_diagnostics} we show the time evolution of the long-range coherence $g^{(1)}(\mathbf r\rightarrow\infty ,t)$, which quantifies the build up of phase coherence from initial phase disorder. At the final time, $t = 60$, the first-order coherence function $g^{(1)}(x)$ along the $x$ direction remains close to unity across the entire system size, confirming the presence of long-range phase coherence characteristic of a superfluid. No density modulation is observed in this regime, which is in accordance with the linear stability analysis predicting the absence of dynamical instabilities for small $\alpha$.

\subsection{Supersolid phase}
\begin{figure}
\centering 
\includegraphics[scale=0.7]{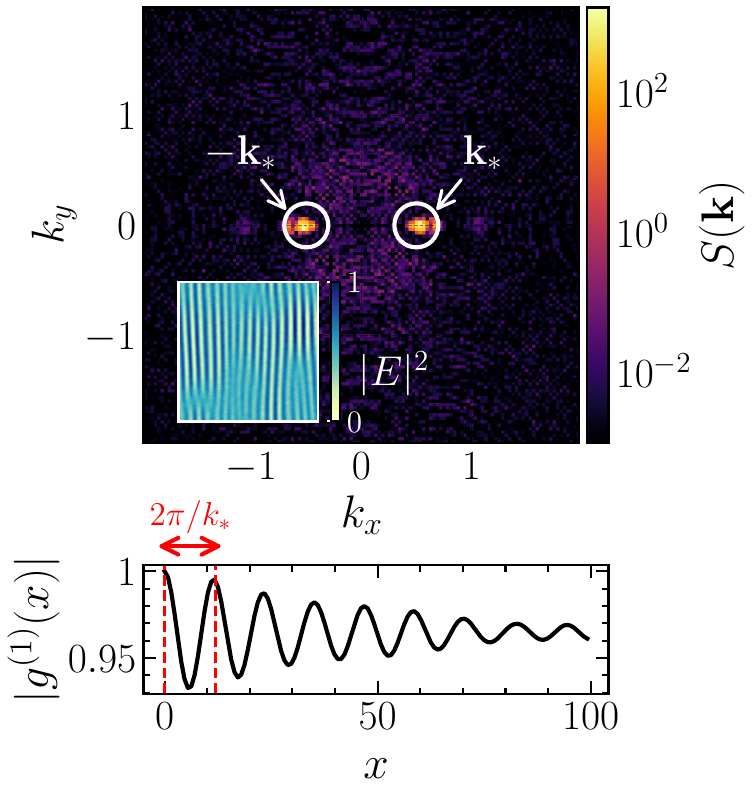}
\caption{Characterization of the supersolid phase in steady state. Top panel: strucrture factor $S(\mathbf k)$ in log scale showing Bragg peaks at finite wavevectors, indicating crystalline order. Inset: real-space density profile $|E(\mathbf r)|^2$ (axis are those of Fig.~1 of the main text). Botton panel: first-order coherence function $g^{(1)}(x)$ along the $x$ direction, showing long-range phase coherence characteristic of a supersolid. Simulation parameters are $\alpha = 1$, $\kappa = 0.1$, $\Gamma = 0.5$, $k_b = 0.45 k_F$. }
\label{fig_SS_diagnostics}
\end{figure}

Increasing the nonlinear parameter $\alpha$ beyond the critical value determined by the linear stability analysis leads to the emergence of a supersolid phase. In this regime, the photon fluid develops spontaneous density modulations and long-range phase coherence, as evidenced by the behavior of the steady-state first-order coherence function $g^{(1)}(x)$ in Fig.~\ref{fig_SS_diagnostics}. The steady-state structure factor $S(\mathbf k)$ exhibits pronounced Bragg peaks at $\mathbf k = \pm \mathbf k_\ast$, indicating the presence of crystalline order, while the density profile $|E(\mathbf r)|^2$ (inset of top panel) shows a periodic modulation consistent with the wavevector of the roton instability predicted by the linear stability analysis. First-order coherence $g^{(1)}(x)$ along the $x$ direction becomes finite over long distances, confirming the coexistence of superfluidity and crystalline order characteristic of a supersolid phase.

\subsection{Solid phase}
\begin{figure}
\centering 
\includegraphics[scale=0.7]{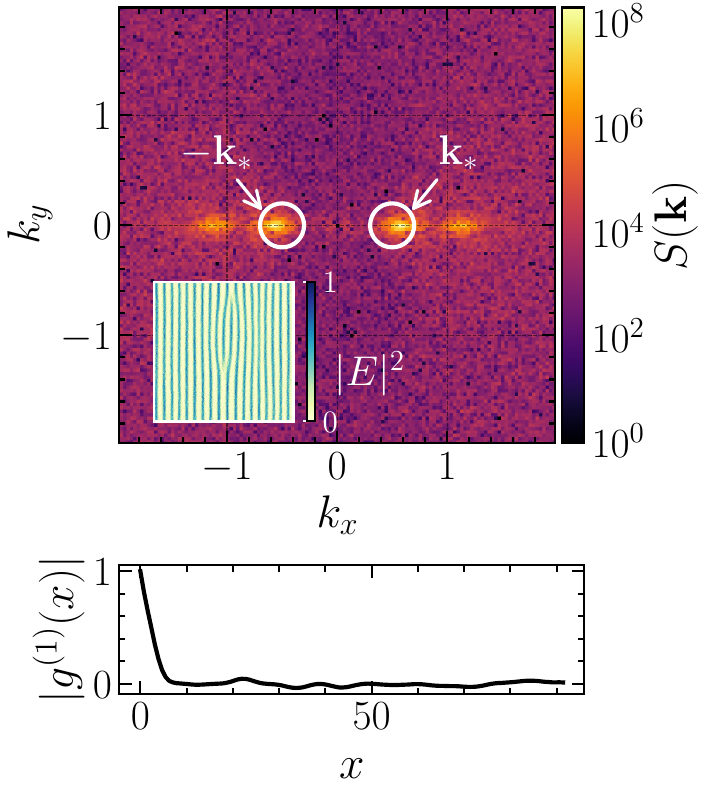}
\caption{Same as Fig.~\ref{fig_SS_diagnostics} but for solid steady-state. Simulation parameters: $\alpha = 10$, $\kappa = 0.1$, $\Gamma = 0.5$, $k_b = 0.45 k_F$}
\label{fig_S_diagnostics}
\end{figure}

Further increasing the nonlinearity $\alpha$ leads to a transition into a solid phase, where the photon fluid exhibits pronounced density modulations and a loss of long-range phase coherence. In this regime, the steady-state first-order coherence function $g^{(1)}(x)$ decays rapidly with distance, indicating the absence of global phase coherence, as depicted in Fig.~\ref{fig_S_diagnostics}, while the structure factor $S(\mathbf k)$ retains sharp Bragg peaks at finite wavevectors, reflecting the strong crystalline order in the density profile $|E(\mathbf r)|^2$. Transition into a solid is not predicted by the linear stability analysis, which only captures the onset of the supersolid phase. The loss of coherence in the solid phase arises from nonlinear effects and interactions beyond the linear regime (see main text).

\section{Experimental set--up design}

\begin{figure*}
\centering
\includegraphics[scale=0.37]{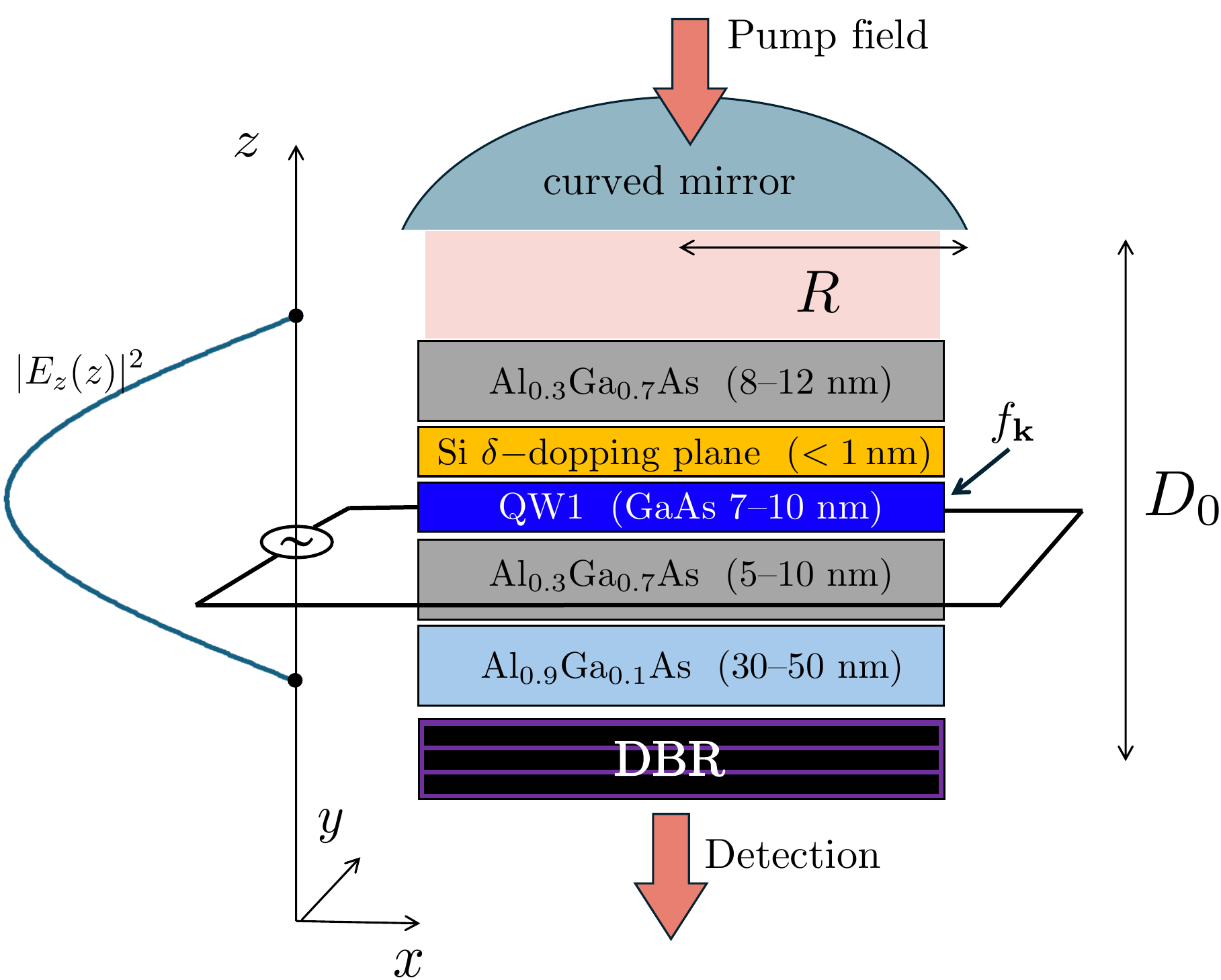}
\caption{Schematic of the proposed cavity architecture used to engineer the momentum-dependent photon--photon interaction. A single GaAs quantum well (QW) is embedded at an antinode of the cavity standing wave, maximizing the photonic sampling of the two-dimensional electron gas (2DEG) and ensuring insensitivity to axial placement errors. Remote Si $\delta$-doping in Al$_{0.3}$Ga$_{0.7}$As at a setback of 8--12 nm supplies an equilibrium sheet density $n_{e}\sim 10^{14}\,\mathrm{m^{-2}}$, placing the electrons well inside the degenerate regime. Lateral electrodes patterned along the in-plane $x$ direction generate a uniform dc field that produces a controlled drift velocity $v_{0}\simeq 0.5\,v_{F}$ in the steady-state electronic distribution. The cavity mode thereby samples a displaced Fermi disk $f_{\mathbf{k}}=f_{0}(\mathbf{k}-m v_{0}\hat{x})$, which induces the asymmetric interaction kernel $g(k_{x},k_{y})$ used in the main text. Pumping is performed from the top mirror with sub-gap light, while detection is collected from the bottom DBR.}
\label{fig_set_up}
\end{figure*}

Figure~\ref{fig_set_up} illustrates the cavity architecture proposed in the main text to engineer the photon--photon interaction required for supersolidity. A planar microcavity supports a standing wave along the growth axis $z$, and a single GaAs quantum well (QW) is placed at an antinode of the intensity profile $|E(z)|^{2}$, maximizing the overlap with the two-dimensional electron gas (2DEG) and making the coupling insensitive to small axial displacements, since $\partial_{z}|E|^{2}=0$ at the antinode. In this configuration the cavity field samples a single degenerate electron sheet, and the resulting interaction kernel is directly proportional to the static density response of the 2DEG,
\begin{equation}
  g(\mathbf k)\propto \chi_0(\mathbf k),
\end{equation}
so that the cavity inherits the full momentum dependence of the electronic Lindhard function.

The embedded heterostructure consists of a GaAs QW of thickness $7$--$10~\mathrm{nm}$, bounded by Al$_{0.3}$Ga$_{0.7}$As barriers and remotely doped by a thin Si $\delta$-layer at a setback $d\simeq 8$--$12~\mathrm{nm}$. For donor densities $N_{D}\sim (1$--$2)\times 10^{16}\,\mathrm{m^{-2}}$, the equilibrium sheet density stabilizes at $n_{e}\sim 10^{14}\,\mathrm{m^{-2}}$, placing the electrons well inside the degenerate regime at $T\simeq 20$--$40~\mathrm{K}$. The resulting sharp Fermi surface is essential to generate the nonlocal structure of $\chi(\mathbf k)$ that produces the roton-like minimum in the photonic dispersion. A transparent top gate (ITO on Al$_{2}$O$_{3}$) patterned on a shallow mesa allows \emph{in situ} tuning of $n_{e}$ without adding significant optical loss; ohmic contacts are placed at the mesa perimeter, outside the field maximum. The cavity is pumped from the top DBR with a weakly focused ($20$--$30~\mu\mathrm{m}$) beam near the $k\simeq 0$ resonance and at sub-gap photon energies, so that interband absorption and hole dynamics remain negligible, and emission is collected from the bottom DBR. Under these conditions the electronic distribution is stationary and thermalized, and the cavity samples a fixed interaction kernel $g(\mathbf k)$ determined solely by the equilibrium 2DEG response.

To generate an anisotropic electronic distribution with controlled drift, the QW mesa incorporates a pair of lateral electrodes separated by $L\simeq 10$--$30~\mu\mathrm{m}$. A small dc bias $\Delta V$ creates an in-plane electric field $E_{x}=\Delta V/L$, which in the linear-response regime produces a steady drift velocity $v_{0}=\mu E_{x}$, with $\mu$ the low-temperature mobility. For $n_{e}\sim 10^{14}\,\mathrm{m^{-2}}$ and GaAs effective mass $m=0.067\,m_{e}$, the Fermi wavevector is $k_{F}=\sqrt{2\pi n_{e}}$ and
\begin{equation}
  v_{F}=\frac{\hbar k_{F}}{m}\simeq 4.5\times 10^{5}\,\mathrm{m/s}.
\end{equation}
Targeting $v_{0}\simeq 0.5\,v_{F}\approx 2.2\times 10^{5}\,\mathrm{m/s}$ and taking a conservative mobility $\mu=1\,\mathrm{m^{2}/(V\,s)}$ yields
\begin{equation}
  E_{x}\simeq \frac{v_{0}}{\mu}\simeq 2.2\times 10^{5}\,\mathrm{V/m},
\end{equation}
so that for $L=20\,\mu\mathrm{m}$ the required bias is $\Delta V\simeq 4.4\,\mathrm{V}$, within the standard operating range of GaAs QW devices. The resulting displaced Fermi disk $f_{\mathbf{k}}=f_{0}(\mathbf{k}-m v_{0}\hat{x})$ is stable on photonic timescales and generates the asymmetric interaction kernel $g(k_{x},k_{y})$ used in the main text.

Finally, the parameter regime considered here corresponds unequivocally to weak light--matter coupling, in contrast with microcavity platforms exhibiting strong coupling and polariton formation~\cite{smolka2014cavity}. For a cavity thickness $D_{0} = 100~\mu\mathrm{m}$ and GaAs refractive index $n\sim 3$, the first longitudinal mode has energy $\hbar\omega_0 \approx 1$--$3~\mathrm{meV}$, while the GaAs band gap is $E_{g} \approx 1.30~\mathrm{eV}$ and typical intersubband transitions lie at $10$--$30~\mathrm{meV}$. The cavity mode is therefore detuned by at least one order of magnitude from any sharp electronic resonance. In this off-resonant regime, the effective coupling between the cavity photon and a transition with bare vacuum Rabi frequency $\Omega_{\mathrm{R}}$ is suppressed as $\Omega_{\mathrm{R,eff}} \sim \Omega_{\mathrm{R}}^{2}/\Delta$, where $\Delta = |\hbar\omega_{\mathrm{cav}} - E_{\mathrm{res}}|$ is the detuning between the cavity photon energy and the electronic transition energy. For $\Delta$ of order $1~\mathrm{eV}$ (interband), one has $\Omega_{\mathrm{R,eff}}\ll \gamma_{\mathrm{cav}} \sim 0.1$--$1~\mathrm{meV}$. The usual strong-coupling criterion $\Omega_{\mathrm{R}} > (\gamma_{\mathrm{cav}} + \gamma_{x})/4$, with $\gamma_{x}$ the electronic linewidth, is thus never satisfied. Thus no vacuum Rabi splitting or anticrossing can occur, and the intracavity field remains a pure photon mode. The 2DEG therefore acts as a dispersive, weakly absorbing medium whose equilibrium susceptibility $\chi(\mathbf{k},0)$ mediates the nonlocal photon--photon interaction kernel $g(\mathbf{k})$ employed throughout this work.

\bibliographystyle{apsrev4-1}
\bibliography{references}

\end{document}